\documentclass[iop]{emulateapj}
\slugcomment{Accepted for publication in ApJ}

\shorttitle{ALMA [CII] Line and Dust Continuum Observation of a $z=7$ LAE}
\shortauthors{Ota and Walter et al.}

\begin{document}


%\title{ALMA Constraint on Ionized Carbon and Dust in a $z=7$ Star-forming Galaxy}
%\title{ALMA Constraint on [CII] and Dust in a $z=7$ Star-forming Galaxy in Reionization Epoch}
%\title{ALMA Constraint on Ionized Carbon and Dust in a $z=7$ Star-forming Galaxy at the Epoch of Reionization}
%\title{ALMA Constraint on Ionized Carbon 158 $\mu$m [CII] and Dust Continuum Emissions of a $z=7$ Normally Star-forming Galaxy in the Epoch of Reionization}
\title{ALMA Observation of 158 $\mu$m [CII] Line and Dust Continuum of a $z=7$ Normally Star-forming Galaxy in the Epoch of Reionization\altaffilmark{*}}
%\thanks{{\it Herschel} is an ESA space observatory with science instruments provided by European-led Principal Investigator consortia and with important participation from NASA.}
%\title{ALMA Constraint on Ionized Carbon and Dust Emissions in a $z \simeq 7$ Ly$\alpha$ Emitter}
%\title{Limits on the [CII] and Dust Continuum Emissions of A $z\sim7$ Ly$\alpha$ Emitter Probed by the ALMA} 


%\author{Kazuaki Ota\altaffilmark{1,2,9}, Fabian Walter\altaffilmark{3}, Kouji Ohta\altaffilmark{4}, Bunyo Hatsukade\altaffilmark{5}, Elisabete da Cunha\altaffilmark{3}, Jorge Gonz\'alez-L\'opez\altaffilmark{6,3}, Roberto Decarli\altaffilmark{3}, Hiroshi Nagai\altaffilmark{5}, Masanori Iye\altaffilmark{5}, Nobunari Kashikawa\altaffilmark{5} et al.}

\author{Kazuaki Ota\altaffilmark{1,2,16}, Fabian Walter\altaffilmark{3}, Kouji Ohta\altaffilmark{4}, Bunyo Hatsukade\altaffilmark{5}, Chris L. Carilli\altaffilmark{6,2}, Elisabete da Cunha\altaffilmark{3}, Jorge Gonz\'alez-L\'opez\altaffilmark{7,3}, Roberto Decarli\altaffilmark{3}, Jacqueline A. Hodge\altaffilmark{3,8}, Hiroshi Nagai\altaffilmark{5}, Eiichi Egami\altaffilmark{9}, Linhua Jiang\altaffilmark{10}, Masanori Iye\altaffilmark{5,11}, Nobunari Kashikawa\altaffilmark{5,11}, Dominik A. Riechers\altaffilmark{12}, Frank Bertoldi\altaffilmark{13}, Pierre Cox\altaffilmark{14,15}, Roberto Neri\altaffilmark{14}, Axel Weiss\altaffilmark{16}}

\affil{$^1$Kavli Institute for Cosmology, University of Cambridge, Madingley Road, Cambridge, CB3 0HA, UK; kota@ast.cam.ac.uk}
\affil{$^2$Cavendish Laboratory, University of Cambridge, 19 J.J. Thomson Avenue, Cambridge, CB3 0HE, UK}
\affil{$^3$Max-Planck Institut f\"ur Astronomie, K\"onigstuhl 17, D-69117, Heidelberg, Germany}
\affil{$^4$Department of Astronomy, Kyoto University, Kitashirakawa-Oiwake-cho, Sakyo-ku, Kyoto 606-8502, Japan}
\affil{$^5$National Astronomical Observatory of Japan, 2-21-1 Osawa, Mitaka, Tokyo, 181-8588, Japan}
\affil{$^6$National Radio Astronomy Observatory, P.O. Box 0, Socorro, NM 87801-0387, USA}
\affil{$^7$Instituto de Astrof\'isica, Facultad de F\'isica, Pontificia Universidad Cat\'olica de Chile, Av.~Vicu\~na Mackenna 4860, 782-0436 Macul, Santiago, Chile}
\affil{$^8$National Radio Astronomy Observatory, 520 Edgemont Road, Charlottesville, VA 22903, USA}
\affil{$^9$Steward Observatory, University of Arizona, 933 North Cherry Avenue, Tucson, AZ 85721, USA}
\affil{$^{10}$School of Earth and Space Exploration, Arizona State University, Tempe, AZ 85287-1504, USA}
\affil{$^{11}$The Graduate University for Advanced Studies, 2-21-1 Osawa, Mitaka, Tokyo, 181-8588, Japan}
\affil{$^{12}$Department of Astronomy, Cornell University, 220 Space Sciences Building, Ithaca, NY 14853, USA}
\affil{$^{13}$Argelander Institute for Astronomy, University of Bonn, Auf dem H\"ugel 71, D-53121 Bonn, Germany}
%\affil{$^{13}$IRAM, 300 rue de la piscine, F-38406 Saint-Martin d'H$\grave{{\rm e}}$res, France}
\affil{$^{14}$Institut de Radio Astronomie Millim\'etrique (IRAM), 300 rue de la piscine, F-38406 Saint-Martin d'H$\grave{{\rm e}}$res, France}
\affil{$^{15}$ALMA SCO, Alonso de Cordova 3107, Vitacura, Santiago, Chile}
\affil{$^{16}$Max-Planck-Institut f\"ur Radioastronomie, Auf dem H\"ugel 69, D-53121 Bonn, Germany}

\altaffiltext{}{---------------------------------------------------------}
\altaffiltext{*}{Based in part on data collected at Subaru Telescope, which is operated by the National Astronomical Observatory of Japan, observations made with the NASA/ESA {\it Hubble Space Telescope}, obtained from the Data Archive at the Space Telescope Science Institute, which is operated by the Association of Universities for Research in Astronomy, Inc.~under NASA contract NAS 5-26555 and observations made with the {\it Spitzer Space Telescope}, which is operated by the Jet Propulsion Laboratory, California Institute of Technology under a contract with NASA. {\it Herschel} is an ESA space observatory with science instruments provided by European-led Principal Investigator consortia and with important participation from NASA.}
\altaffiltext{16}{Kavli Institute Fellow}
%\altaffiltext{17}{Hubble Fellow}

%\altaffiltext{5}{Department of Astronomy, Graduate School of Science, University of Tokyo, 7-3-1 Hongo, Bunkyo-ku, Tokyo 113-0033, Japan}


\begin{abstract}
We present ALMA observations of the [CII] line and far-infrared (FIR) continuum of a normally star-forming galaxy in the reionization epoch, the $z=6.96$ Ly$\alpha$ emitter (LAE) IOK-1. Probing to sensitivities of $\sigma_{\rm line} =$ 240 $\mu$Jy beam$^{-1}$ (40 km s$^{-1}$ channel) and $\sigma_{\rm cont} =$ 21 $\mu$Jy beam$^{-1}$, we found the galaxy undetected in both [CII] and continuum. Comparison of UV--FIR spectral energy distribution (SED) of IOK-1, including our ALMA limit, with those of several types of local galaxies (including the effects of the cosmic microwave background, CMB, on the FIR continuum) suggests that IOK-1 is similar to local dwarf/irregular galaxies in SED shape rather than highly dusty/obscured galaxies. Moreover, our $3 \sigma$ FIR continuum limit, corrected for CMB effects, implies intrinsic dust mass $M_{\rm dust} < 6.4 \times 10^7 M_{\odot}$, FIR luminosity $L_{\rm FIR} < 3.7\times 10^{10} L_{\odot}$ (42.5--122.5 $\mu$m), total IR luminosity $L_{\rm IR} < 5.7 \times 10^{10} L_{\odot}$ (8--1000 $\mu$m) and dust-obscured star formation rate (SFR) $< 10$ $M_{\odot}$ yr$^{-1}$, if we assume that IOK-1 has a dust temperature and emissivity index typical of local dwarf galaxies. This SFR is 2.4 times lower than one estimated from the UV continuum, suggesting that $< 29$\% of the star formation is obscured by dust. Meanwhile, our $3 \sigma$ [CII] flux limit translates into [CII] luminosity, $L_{\rm [CII]} < 3.4 \times 10^7 L_{\odot}$. Locations of IOK-1 and previously observed LAEs on the $L_{\rm [CII]}$ vs.~SFR and $L_{\rm [CII]}/L_{\rm FIR}$ vs.~$L_{\rm FIR}$ diagrams imply that LAEs in the reionization epoch have significantly lower gas and dust enrichment than AGN-powered systems and starbursts at similar/lower redshifts, as well as local star-forming galaxies. 
\end{abstract}





\section{Introduction}
%While such properties of low-$z$ ($z<3$) star-forming galaxies have been well studied using the molecular CO lines (and far-infrared (FIR) continuum) as a tracers, those of high-$z$ star-forming galaxies have not due to its faintness. The ionized carbon 158 micron [CII] line is a powerful alternative to detect such high-$z$ galaxies, because it is the strongest cooling line of interstellar medium (ISM).
Probing visible and dust-obscured star formation and physical properties of galaxies from the local to the early Universe, including the epoch of cosmic reionization at redshift $z > 6$, is key to understanding the formation and evolution of galaxies, roles of galaxies in reionization and their relation among diverse galaxy populations. To date, objects such as quasars, active galactic nuclei (AGNs), their host galaxies, submm-detected starburst galaxies (SMGs), normal star-forming galaxies and (ultra)luminous infrared galaxies ((U)LIRGs) have been detected at rest frame far-infrared (FIR) wavelengths in the local to high redshift Universe \citep[e.g.,][]{Dunne00,Malhotra01,Boselli02,Iono06,Maiolino09,Ivison10,Stacey10,Wagg10,Cox11,DeBreuck11,DeLooze11,Skibba11,Venemans12,Wang13,Willott13,Walter12a,Carilli13,Riechers13}. The CO rotational transition lines have been often used as a tracer of molecular gas to probe both visible and obscured SFRs, properties of molecular gas clouds and dynamics of galaxies. However, at $z > 6$, detections of the CO line have been limited to only the quasars/AGNs, their hosts and an SMG \citep[e.g.,][]{Walter03,Wang10,Riechers13}, because highly excited CO lines only seen in these objects are bright, while the CO lines in $z > 6$ normally star-forming galaxies are likely not as highly excited and thus too faint to detect \citep[e.g.,][]{Wagg09,Wagg12a}. Although SFRs of $z > 6$ normally star-forming galaxies such as Ly$\alpha$ emitters (LAEs) and Lyman break galaxies (and related properties such as Ly$\alpha$/UV luminosity function/density) are also an important probe of reionization, their dust-obscured SFRs and other relevant physical properties have been still poorly constrained \citep[e.g.,][]{Hu10,Ota10a,Ouchi10,Kashikawa11,Bouwens12,Shibuya12,Finkelstein12,Finkelstein13,Dunlop13,Robertson13,Momose14,Konno14}.

Alternatively, the ionized carbon 157.74 $\mu$m [CII] line (rest frame frequency 1900.54 GHz) has been observed to be the strongest cooling line of interstellar medium (ISM) in many galaxies \citep[e.g., a few hundred to a few thousand times stronger than CO(1--0), CO(2--1), CO(3--2) and CO(4--3) lines in $z\sim 1$--2 star-formation/AGN-dominated galaxies,][]{Stacey10}, and its luminosity is as high as $\sim$ 0.01--1\% of total FIR luminosity \citep[e.g.,][]{Maiolino05,Maiolino09,Iono06,Ivison10,Stacey10,Wagg10,DeBreuck11}. When redshifted to $z > 6$, the [CII] line is observable in millimeter wavelengths from the ground. Even so, to date, only a handful number of quasar/AGN host galaxies and an SMG have been detected in [CII] at $z > 6$ \citep[e.g.,][]{Maiolino05,Venemans12,Wang13,Willott13,Riechers13}, and there is no detection in normally star-forming galaxies in the reionization epoch \citep{Boone07,Walter12b,Gonzalez-Lopez14} even though it is thought to be the strongest line in FIR regime in these galaxies. However, the advent of the Atacama Large Millimeter/submillimeter Array (ALMA) has the potential to change this situation as it has an order of magnitude higher sensivity than any other current millimeter interferometers. Even in its early science operation, it has been argued that ALMA could potentially detect the [CII] lines of normally star-forming galaxies in the reionization epoch at $z > 6$ in only a few hours and resolve them on a few kpc scales. Motivated by this, we conducted a [CII] line and FIR continuum observation of such a galaxy with ALMA.
%When redshifted to $z > 6$, the [CII] line is observable at millimeter wavelengths from the ground.

Among many $z > 6$ star-forming galaxies found to date, we chose to observe a $z = 6.96$ LAE that we had discovered, IOK-1 \citep{Iye06,Ota08}, because it is one of the handful of spectroscopically confirmed highest-class redshift ($z \gtrsim 7$) objects, and we have a wealth of ancillary data availaible for this galaxy that were analyzed in the context of the ALMA observation. For example, we have conducted IOK-1's photometry and spectroscopy at optical to mid-infrared wavelengths and derived physical properties such as Ly$\alpha$ and UV luminosities/SFRs, stellar mass and morphology at UV wavelengths \citep{Ota08,Ota10b,Cai11,Jiang13}. In the present paper, we investigate the properties of IOK-1 in the [CII] line and FIR continuum. The paper is organized as follows. In \textsection 2, we describe our ALMA observation of IOK-1. Then, in \textsection 3, we analyze the properties of IOK-1 in FIR continuum and [CII] line such as SED, effects of cosmic microwave background (CMB) on dust continuum, dust mass, infrared (IR) luminosity and dust-obscured SFR as well as the relation between SFR and [CII] luminosity of high redshift LAEs and [CII] to FIR luminosity ratio, and compare these properties with those of other galaxy populations in the local to high redshift Universe, and discuss its implication for galaxy formation and evolution. We summarize and conclude our study in \textsection 4. Throughout, we adapt a concordance cosmology with $(\Omega_m, \Omega_{\Lambda}, h)=(0.3, 0.7, 0.7)$.  
%Throughout, we adapt a concordance cosmology with $(\Omega_m, \Omega_{\Lambda}, h)=(0.3, 0.7, 0.7)$ consistent with the recent CMB constraints (Hinshaw et al.~2013; Planck Collaboration~2013).  

%%figure 1

\begin{figure}
\epsscale{1.17}
\plotone{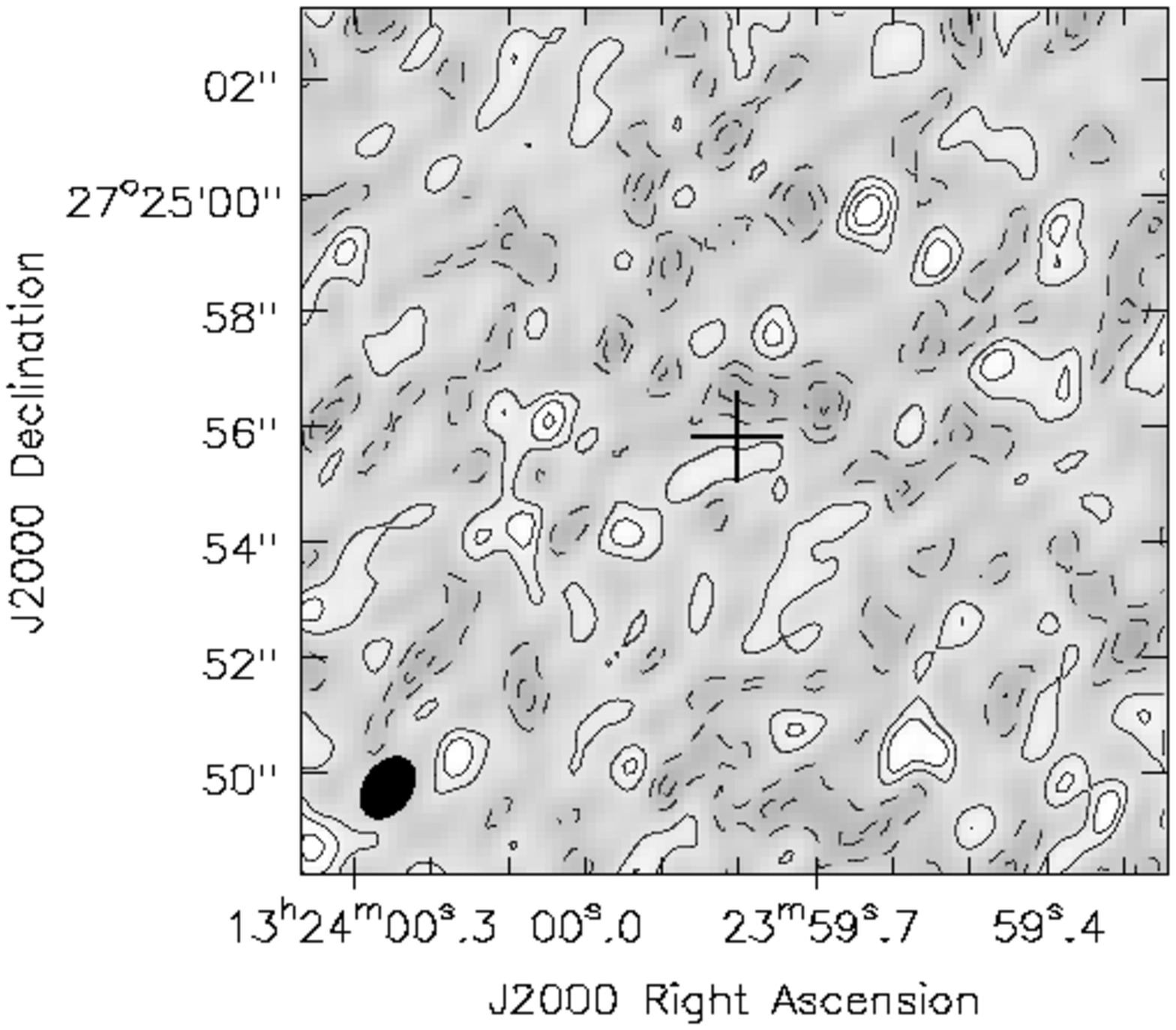}
\caption{The rest frame 158 $\mu$m FIR continuum image of IOK-1 at a resolution of $1\farcs1 \times 0\farcs75$. Coordinates are in the J2000.0 system and contours are in steps of $1\sigma =$ 18 $\mu$Jy beam$^{-1}$, starting at $\pm 1\sigma$. Positive and negative contours are presented with solid and dashed lines, respectively. The synthesized beam is shown in the bottom left. The position of IOK-1 measured in the optical narrowband NB973 (Ly$\alpha$ line plus rest frame UV continuum) image \citep{Iye06,Ota08} is marked with the cross. No emission is detected. \label{FIR_Continuum_IOK1}}
\end{figure}
\vspace*{0.5cm}
%\vspace*{1cm}

%%figure 2

\begin{figure*}
\epsscale{1.17}
%\plottwo{Figure_SEDs_1.eps}{Figure_SEDs_Spirals_Dwarfs.eps}
%\plottwo{Figure_SEDs_1_w_PdBI_CARMA_IOK-1limits.eps}{Figure_SEDs_Spirals_Dwarfs_w_PdBI_CARMA_IOK-1limits.eps}
%\plottwo{Figure_SEDs_1_w_PdBI_CARMA_IOK-1limits_NIRinset-II.eps}{Figure_SEDs_Spirals_Dwarfs_w_PdBI_CARMA_IOK-1limits_NIRinset-II.eps}
\plottwo{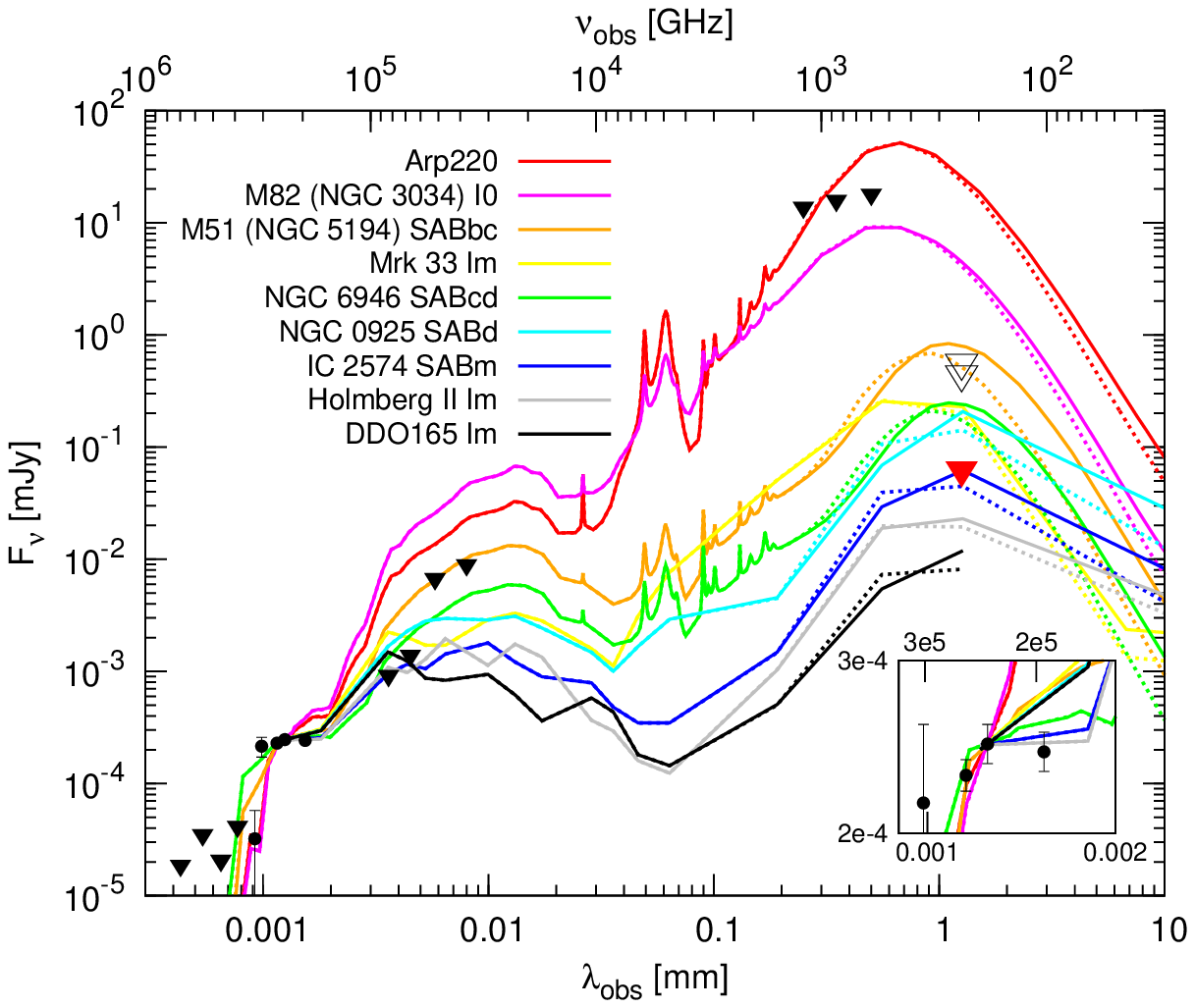}{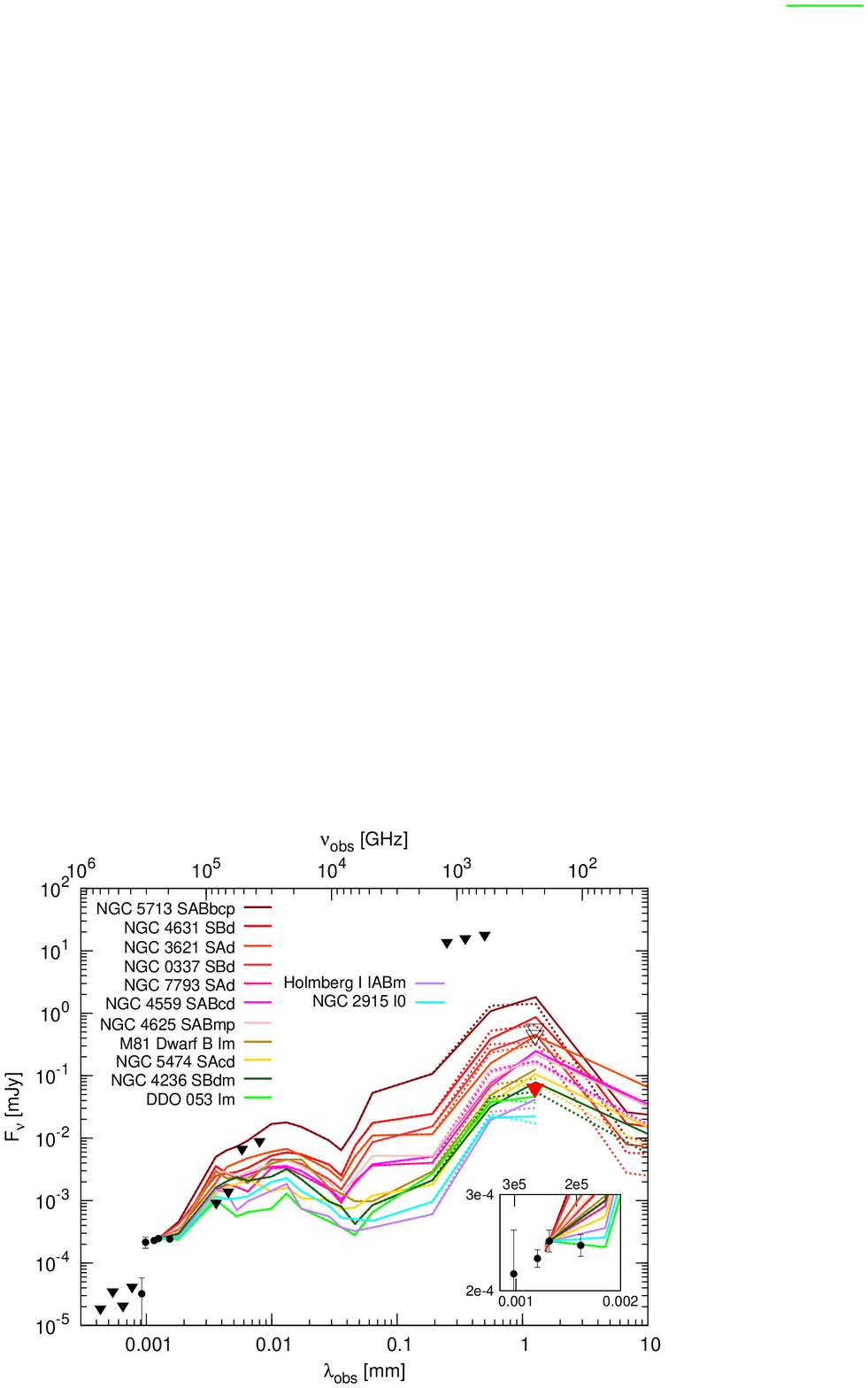}
\caption{The SED of IOK-1 in the observer's frame (bottom axis: wavelength, top axis: frequency). Detections in near-infrared wavelengths are shown by the black circles: Subaru Telescope/Suprime-Cam $z'$ and $y$ bands \citep{Ota10b} and {\it HST}/WFC3 F110W, F125W and F160W bands \citep{Cai11,Jiang13}. The $3\sigma$ upper limits on the flux densities in optical to millimeter wavelengths are indicated by the black triangles: Subaru/Suprime-Cam $B$, $V$, $R$ and $i'$ bands \citep{Ota08}, {\it Spitzer}/IRAC 3.6, 4.5, 5.8 and 8.0 $\mu$m bands \citep{Ota10b} and {\it Herschel}/SPIRE 250, 350 and 500 $\mu$m bands \citep[][Ivison, Eales \& Smith 2013 private communication]{Eales10,Pascale11}. Our ALMA $3\sigma$ upper limit on 1.26 mm (rest frame 158 $\mu$m) continuum is shown by the red triangle. For comparison, we also plot $3\sigma$ upper limits on 1.26 mm continuum from the previous IRAM/PdBI and CARMA observations of IOK-1 \citep[lower and upper open triangles, respectively;][]{Walter12b,Gonzalez-Lopez14}. Several different types of local galaxy SEDs are shown by the color coded lines \citep[See the labels in the figure. The galaxy morphological types were taken from][]{Dale07} for comparison. They were redshifted to $z=6.96$ and normalized to the near-infrared (rest frame UV) flux of IOK-1 redward of Ly$\alpha$ (F125W band flux). The solid lines are the original SEDs with their submm--mm continua (cold dust emission) not including any effects of CMB, while the dotted lines with the same color coding are the SEDs including the effects of CMB (see sections \textsection \ref{CMBeffect} and \ref{SED} for details). (Left panel) Comparison with early, late, dwarf and irregular types of galaxies. The SEDs of Arp 220, M82, M51 and NGC 6946 were taken from \citet{Silva98}, while those of Mrk 33, NGC 0925, IC 2574, Holmberg II and DDO 165 were from \citet{Dale07}. (Right panel) Comparison with spiral/barred, dwarf spiral, dwarf and irregular types of galaxies, with SEDs taken from \citet{Dale07}. The insets show close-ups of the plots around the near-infrared (rest frame UV) continuum including $y$, F110W, F125W and F160W band data points.\label{SED_IOK-1}}
\end{figure*}

\section{Observation}
We observed IOK-1 by targeting its [CII] line redshifted to $z=6.96$ measured from Ly$\alpha$ emission (1.26 mm, 238.76 GHz in the observer's frame or 157.74 $\mu$m, 1900.54 GHz at rest frame) and the underlying dust continuum emission. We used the ALMA band 6 with the Extended Array consisting of 19--24 12-m diameter antennas on 15 June, 15 and 27 July and 29 November 2012 during the Cycle 0 period. We set the two spectral windows (SPWs) in the lower sideband to the continuum at the lower frequency side of the [CII] line, and the third and fourth SPWs in the upper sideband to the redshifted [CII] line and the continuum  at the higher frequency side, respectively, in order to cover the continua at the both sides of [CII] line. Specifically, we set widths of all the four SPWs to 2 GHz (1.875 GHz after discarding edge channels) using the Time Domain Mode (TDM) for dual polarization with central frequencies of 223.76, 225.76, 238.76 and 240.76 GHz. The channel spacing of the TDM mode is 15.6 MHz, and this gives a spectral resolution of 40 km s$^{-1}$. We obtained 2.7 hours of on-source integration. We carried out data reduction with the Common Astronomy Software Applications (CASA). The flux density was scaled with the observation data of 3C279 (J1256-057) for one of the four execution blocks and Titan for the rest of three execution blocks. The typical flux uncertainties in the millimeter regime are $\sim 10$\%. J1310+323 was used for phase calibration. 3C279 was also used for bandpass calibration. Finally, we imaged the [CII] line and FIR continuum, using `natural' weighting, and reaching the sensitivities of $\sigma_{\rm line} =$ 215 $\mu$Jy beam$^{-1}$ per TDM channel (40 km s$^{-1}$) and $\sigma_{\rm cont} =$ 18 $\mu$Jy beam$^{-1}$ for 7.5 GHz bandwidth (4 SPWs $\times$ 1.875 GHz), respectively. The resolutions (synthesized beam sizes) of [CII] line and FIR continuum images presented here are $1\farcs1 \times 0\farcs75$.

%We obtain a flux density of XX Jy for our gain calibrator, XXX, located XX from IOK-1 at the time of observation. (Add here description about weather conditions and humidity herer?) (Add here information about system temperature $T_{\rm sys}$, water vapor radiometer correction here?) (Add uncertainty in flux here?) (Write about self-calibration here?).

%% In this section, we use  the \subsection command to set off
%% a subsection.  \footnote is used to insert a footnote to the text.

\section{Result and Discussion}
\subsection{FIR Continuum Properties \label{FIRcont_Prop}}
Figure \ref{FIR_Continuum_IOK1} shows the FIR continuum map around IOK-1 at $1\farcs1 \times 0\farcs75$ resolution. The galaxy is not significantly detected. IOK-1 has a size of $\sim 1\farcs6 \times 1\farcs0$ in the Subaru Telescope Ly$\alpha$ narrowband NB973 image \citep{Iye06,Ota08}, which is larger than the sizes in any other images (rest frame UV continuum) where IOK-1 is detected. Hence, we convolved the FIR continuum map to the resolution of $1\farcs5 \times 1\farcs2$ (through tapering with a 100 k$\lambda$ taper) to see if we detect any extended source. However, we did not detect IOK-1 in this map, either. As the rms of this map is $\sigma_{\rm cont}=21$ $\mu$Jy beam$^{-1}$, the $3\sigma$ upper limit on 1.26 mm FIR continuum of IOK-1 is $F_{\nu/(1+z)}^{\rm obs} < 63$ $\mu$Jy. In this section, based on this flux limit, we constrain the dust properties of IOK-1 such as its possible SED shape, dust mass, total IR luminosity and dust-obscured SFR by considering the effects of cosmic microwave background (CMB) on the observed continuum flux. 
%The $3\sigma$ upper limit on 1.26 mm FIR continuum per synthesized beam is 54 $\mu$Jy.
%This is about twice as large as the ALMA beam size $1\farcs1 \times 0\farcs75$. Hence, the $3\sigma$ upper limit on 1.26 mm FIR continuum of IOK-1 is $\sim \sqrt{2}$ times worse than one per beam, which is 

\subsubsection{Effects of CMB on FIR Continuum Measurements \label{CMBeffect}}
The study by \citet{daCunha13} describes how the CMB affects measurements of FIR continuum emission of high redshift galaxies. Hence, before we derive and discuss constraints on the dust properties of IOK-1 from the observed 1.26 mm continuum flux limit, we first describe the effects of CMB on FIR continuum and how we actually apply them to our observed continuum flux limit \citep[see also][for details]{Gonzalez-Lopez14}. 

CMB affects measurements of FIR continuum in two ways. One is that CMB heats the dust of galaxies to higher temperature as redshift gets higher. According to \citet{daCunha13}, the dust temperature at a redshift $z$ is 
\begin{equation}
T_{\rm dust}(z) = ((T_{\rm dust}^{z=0})^{4+\beta}+(T_{\rm CMB}^{z=0})^{4+\beta}[(1+z)^{4+\beta}-1])^{\frac{1}{4+\beta}}
\label{Tdust}
\end{equation}
Here, $T_{\rm dust}^{z=0}$ is the temperature of dust heated by and in thermal equilibrium with the radiation field produced by stars in a galaxy at $z=0$, where the effects of CMB radiation are negligible. Also, $T_{\rm CMB}^{z=0}=2.73$ K and $\beta$ are temperature of CMB at $z=0$ and dust emissivity index of a thermal spectrum. This equation suggests that effect of CMB heating on dust temperature becomes non-negligible at high redshifts \citep[e.g., see Figure 1 in][for a case of $T_{\rm dust}^{z=0}=18$ K and $\beta=2.0$]{daCunha13}. The ratio (hereafter referred to as $M_{\nu}$) between the dust continuum flux densities at a rest frame frequency $\nu$ and a redshift $z$ when including dust heating by CMB and not including it is
\begin{equation}
M_{\nu} = \frac{F^*_{\nu/(1+z)}}{F^{\rm int}_{\nu/(1+z)}} = \frac{B_{\nu}(T_{\rm dust}(z))}{B_{\nu}(T_{\rm dust}^{z=0})}
\label{M_nu}
\end{equation}
where $F^{\rm int}_{\nu/(1+z)}$, $F^*_{\nu/(1+z)}$ and $B_{\nu}$ are the intrinsic dust continuum flux density of a galaxy, the dust continuum flux density after including the effect of CMB heating of dust and the Planck function. This $M_{\nu}$ is a factor for correcting the effect of CMB heating of dust \citep[for example, see Figure 3 top panel of][for the case of $T_{\rm dust}^{z=0}=18$ K and $\beta=2$]{daCunha13}. 

Meanwhile, another effect of CMB on dust continuum measurements is that CMB itself becomes a background against which FIR continuum flux should be measured. \citet{daCunha13} have shown that the ratio (hereafter referred to as $C_{\nu}$) between the flux at a frequency $\nu$ that can be measured against CMB at a redshift $z$ and the flux emitted by dust in a galaxy at that frequency and redshift including heating of dust by CMB is
\begin{equation}
C_{\nu} = 1 - \frac{B_{\nu}(T_{\rm CMB}(z))}{B_{\nu}(T_{\rm dust}(z))}
\label{C_nu}
\end{equation}
where $T_{\rm CMB}(z) = T_{\rm CMB}^{z=0} (1+z)$ is a temperature of CMB at a redshift $z$. This is a factor for correcting the effect of CMB as background \citep[for example, see Figure 3 bottom panel of][for the case of $T_{\rm dust}^{z=0}=18$ K and $\beta=2$]{daCunha13}. 

Finally, we can estimate the intrinsic FIR continuum flux $F_{\nu/(1+z)}^{\rm int}$ emitted by a galaxy from the observed continuum flux $F_{\nu/(1+z)}^{\rm obs}$ by using the equation
\begin{eqnarray}
F_{\nu/(1+z)}^{\rm obs} &=& C_{\nu} \times M_{\nu} \times F_{\nu/(1+z)}^{\rm int} \nonumber \\
                        &=& \left[ \frac{B_{\nu}(T_{\rm dust}(z)) - B_{\nu}(T_{\rm CMB}(z))}{B_{\nu}(T_{\rm dust}^{z=0})} \right] F_{\nu/(1+z)}^{\rm int}
\label{F_int}
\end{eqnarray}
and the equation (1) by assuming $T_{\rm dust}^{z=0}$ and $\beta$.

\begin{figure}
\epsscale{1.2}
\plotone{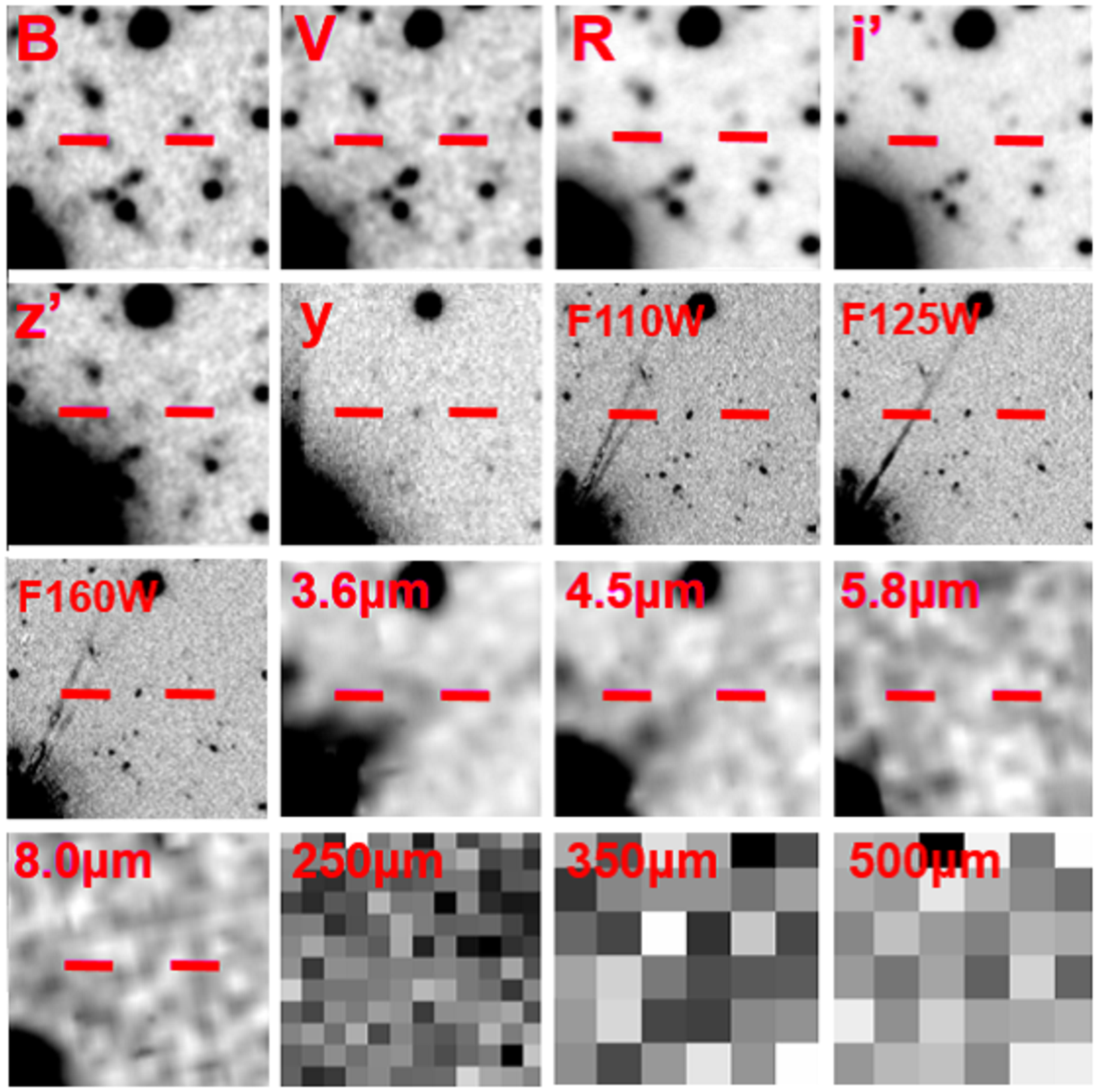}
\caption{The images of IOK-1 at optical \citep[Subaru Telescope Suprime-Cam $BVRi'z'y$ bands:][]{Ota08,Ota10b,Jiang13}, near-infrared \citep[{\it HST} WFC3 F110W, F125W and F160W bands:][]{Cai11,Jiang13}, mid-infrared \citep[{\it Spitzer} IRAC 3.6, 4.5, 5.8 and 8.0 $\mu$m bands:][]{Ota10b} and FIR \citep[{\it Herschel} SPIRE 250, 350 and 500 $\mu$m bands:][Ivison, Eales \& Smith 2013 private communication]{Eales10,Pascale11}. North is up and east to the left. The optical to mid-infrared images are $20'' \times 20''$ and FIR images $60'' \times 60''$ in size. Fluxes measured in these images were used to reproduce the SED of IOK-1 in Figure \ref{SED_IOK-1}. \label{Images_IOK1}}
\end{figure} 

%Table 1 (deluxetable)

\begin{deluxetable*}{cccccccccccc}
\tabletypesize{\scriptsize}
%\rotate
\tablecaption{Summary of the ALMA Observation and Derived Properties of IOK-1\label{ObsProperty}}
%\tablewidth{0pt}
\tablewidth{510pt}
\tablehead{
\colhead{\hspace{-2cm}$z$$^{\rm a}$} & \colhead{\hspace{-3cm}$\nu_{\rm obs}$$^{\rm b}$} & \colhead{\hspace{-1cm}$\sigma_{\rm cont}$$^{\rm c}$} & \colhead{\hspace{-0.1cm}$\sigma_{\rm line}$$^{\rm d}$} & \colhead{$S_{\rm cont}^{\rm obs}$$^{\rm e}$} & \colhead{$S_{\rm line}^{\rm obs}$$^{\rm f}$}
& \colhead{$L_{\rm [CII]}$$^{\rm g}$} & \colhead{$M_{\rm dust}$$^{\rm h}$} & \colhead{$L_{\rm FIR}$$^{\rm h}$} & \colhead{$L_{\rm IR}$$^{\rm h}$} 
& \colhead{SFR$_{\rm dust}$$^{\rm h}$} & \colhead{SFR$_{\rm UV}$$^{\rm i}$}\\
\colhead{\hspace{-2cm}} & \colhead{\hspace{-3cm}(GHz)} & \colhead{\hspace{-1cm}($\mu$Jy beam$^{-1}$)} & \colhead{\hspace{-0.1cm}($\mu$Jy beam$^{-1}$)} & \colhead{($\mu$Jy)} & \colhead{($\mu$Jy)} & \colhead{(10$^7$$L_{\odot}$)} & \colhead{($10^7$$M_{\odot}$)} & \colhead{(10$^{10}$$L_{\odot}$)} & \colhead{(10$^{10}$$L_{\odot}$)} & \colhead{($M_{\odot}$yr$^{-1}$)} & \colhead{($M_{\odot}$yr$^{-1}$)}
}
\startdata
%$z$$^{\rm a}$ & $\nu_{\rm obs}$$^{\rm b}$ & $\sigma_{\rm cont}$$^{\rm c}$ & $\sigma_{\rm line}$$^{\rm d}$ & $S_{\rm cont}^{\rm obs}$$^{\rm e}$ & $S_{\rm line}^{\rm obs}$$^{\rm f}$ & $L_{\rm [CII]}$$^{\rm g}$ & $M_{\rm dust}$$^{\rm h}$ & $L_{\rm FIR}$$^{\rm h}$ & $L_{\rm IR}$$^{\rm h}$ & SFR$_{\rm dust}$$^{\rm h}$ & SFR$_{\rm UV}$$^{\rm i}$\\
%6.96 & (GHz) & ($\mu$Jy beam$^{-1}$) & ($\mu$Jy beam$^{-1}$) & ($\mu$Jy) & ($\mu$Jy) & (10$^7$$L_{\odot}$) & ($10^7$$M_{\odot}$) & (10$^{10}$$L_{\odot}$) & (10$^{10}$$L_{\odot}$) & ($M_{\odot}$yr$^{-1}$) & ($M_{\odot}$yr$^{-1}$)\\\hline
\hspace{-2cm}6.96 & \hspace{-3cm}238.76 & \hspace{-1cm}21 & \hspace{-0.1cm}240 & $<63$ & $<720$ & $<3.4$ & $<6.4$ & $<3.7$ & $<5.7$ & $<10.0$ & 23.9
\enddata
%% Text for table notes should follow after the \enddata but before
%% the \end{deluxetable}. Make sure there is at least one \tablenotemark
%% in the table for each \tablenotetext.
\tablecomments{All the upper limits are $3\sigma$ at a resolution of $1\farcs5 \times 1\farcs2$ (See \textsection \ref{FIRcont_Prop} for details.)}
\tablenotetext{a}{The redshift measured from the Ly$\alpha$ of IOK-1 \citep{Iye06,Ota08}. \citet{Ono12} also reported $z=6.965$.}
\tablenotetext{b}{Observing frequency tuned to the redshift measured from the Ly$\alpha$ of IOK-1.}
\tablenotetext{c}{Continuum sensitivity at 1.26 mm (rest frame 158 $\mu$m) at a resolution of $1\farcs5 \times 1\farcs2$.}
\tablenotetext{d}{[CII] line sensitivity over a channel width of 40 km s$^{-1}$ at a resolution of $1\farcs5 \times 1\farcs2$.}
\tablenotetext{e}{$3\sigma$ continuum flux density limit.}
\tablenotetext{f}{$3\sigma$ [CII] line flux density limit over a channel width of 40 km s$^{-1}$.}
\tablenotetext{g}{[CII] line luminosity limit over a channel width of 40 km s$^{-1}$. We assumed $L_{\rm line}=1.04 \times 10^{-3} F_{\rm line}\nu_{\rm rest}(1+z)^{-1}D_{\rm L}^2$, where the line luminosity, $L_{\rm line}$, is measured in $L_{\odot}$, the velocity-integrated flux, $F_{\rm line}= S_{\rm line}\Delta v$, in Jy km s$^{-1}$ with a velocity width, $\Delta v = 40$ km s$^{-1}$, the rest frequency, $\nu_{\rm rest}=\nu_{\rm obs} (1+z)$, in GHz, and the luminosity distance, $D_{\rm L}$, in Mpc \citep{Walter12b}.}
\tablenotetext{h}{Intrinsic dust mass, FIR luminosity (42.5--122.5 $\mu$m), total IR luminosity (8--1000 $\mu$m) and dust-obscured SFR all corrected for the effects of CMB. These quantities including the effects of CMB are a factor of $C_{\nu}M_{\nu}=0.76$ lower (see \textsection \ref{CMBeffect} and \ref{FIR_Properties} for details).}
\tablenotetext{i}{SFR derived from the UV continuum of IOK-1 by \citet{Jiang13}.}
\end{deluxetable*}

\subsubsection{Spectral Energy Distribution\label{SED}}
Since we have obtained $3\sigma$ upper limit on the observed 1.26 mm continuum flux of IOK-1, $F_{\nu/(1+z)}^{\rm obs} < 63$ $\mu$Jy, we can infer what type of galaxy this object is from its SED at rest frame UV to FIR (observer frame optical to millimeter) including our ALMA flux limit. In Figure \ref{SED_IOK-1}, we compare the SED of IOK-1 with template SEDs of several different types of local galaxies. We reproduce the SED of IOK-1 using the fluxes measured in the optical to FIR images taken with Subaru Telescope Suprime-Cam $BVRi'z'y$ bands \citep{Ota08,Ota10b,Jiang13}, {\it Hubble Space Telescope} ({\it HST}) Wide Field Camera 3 (WFC3) F110W, F125W and F160W bands \citep{Cai11,Jiang13}, {\it Spitzer Space Telescope} Infrared Array Camera (IRAC) 3.6, 4.5, 5.8 and 8.0 $\mu$m bands \citep{Ota10b}, {\it Herschel Space Telescope} Spectral and Photometric Imaging Receiver (SPIRE) 250, 350 and 500 $\mu$m bands \citep[the images of IOK-1 are cut out from the completed {\it Herschel}-ATLAS North Galactic Plane (NGP) project images;][Ivison, Eales \& Smith 2013 private communication]{Eales10,Pascale11} and our ALMA $3\sigma$ upper limit on 1.26 mm dust continuum. These images are shown in Figure \ref{Images_IOK1}. IOK-1 is not detected in any of the SPIRE bands where the beam-convolved map $1\sigma$ point source sensitivities (including both instrumental and confusion noises) are 4.6, 5.3, and 6.0 mJy for 250, 350 and 500 $\mu$m bands, respectively. Note that for comparison, in Figure \ref{SED_IOK-1}, we also plot $3\sigma$ upper limits on 1.26 mm continuum from the previous IRAM/Plateau de Bure Interferometer (PdBI) and Combined Array for Research in Millimeter-wave Astronomy (CARMA) observations of IOK-1 conducted by \citet{Walter12b} and \citet{Gonzalez-Lopez14}, which are shown by the lower and upper open triangles, respectively. Our ALMA upper limit is about one order of magnitude deeper than these previous limits.   

We compare the UV--FIR fluxes/flux limits of IOK-1 with the local galaxy templates of Arp 220, M82, M51, NGC 6946 (the left panel in Figure \ref{SED_IOK-1}) taken from \citet{Silva98} and other local spiral/barred, dwarf and irregular galaxies (Mrk 33, NGC 0925, IC 2574, Holmberg II and DDO 165 in the left panel and all the galaxies in the right panel in Figure \ref{SED_IOK-1}) from \citet{Dale07}. We shift these galaxy SEDs to the redshift of IOK-1, $z=6.96$, normalize them to the rest frame UV continuum flux density redward of Ly$\alpha$ (the {\it HST}/F125W band flux) and plot them by the color coded solid lines in Figure \ref{SED_IOK-1}. We also show their galaxy morphological types in the figure. These are the SEDs not including the effects of CMB on their FIR continua. 

We also plot the SEDs of these template galaxies with their FIR continua (a part of the SED that corresponds to the emission of the cold dust) including the effects of CMB by the color coded dashed lines in Figure \ref{SED_IOK-1}. We convert the original FIR continua of the template galaxies, $F_{\nu/(1+z)}^{\rm int}$, to those including the effects of CMB, $F_{\nu/(1+z)}^{\rm obs}$, by using their dust temperatures $T_{\rm dust}^{z=0}$ and emissivity indices $\beta$ obtained from literature and equations (\ref{Tdust})--(\ref{F_int}) but following the prescription given by \citet{Gonzalez-Lopez14} to calculate the factor $M_{\nu}$ as follows. First, from the literature that conducted single-component modified black body (MBB) fittings to the observed FIR continua of the galaxies, we adopt $T_{\rm dust}^{z=0}=66.7$ K and $\beta=1.83$ for Arp 220 \citep{Rangwala11}, $T_{\rm dust}^{z=0}=48.0$ K and $\beta=1.0$ for M82 \citep{Colbert99}, $T_{\rm dust}^{z=0}=24.9$ K and $\beta=2.0$ for M51 \citep{MentuchCooper12}, $T_{\rm dust}^{z=0}=40.4$ K and $\beta=1.3$ for Mrk 33 \citep{Dunne00}, $T_{\rm dust}^{z=0}=26.0$ K and $\beta=1.5$ for NGC 6946, $T_{\rm dust}^{z=0}=23.7$ K and $\beta=1.5$ for NGC 0925, $T_{\rm dust}^{z=0}=25.9$ K and $\beta=1.5$ for IC 2574, $T_{\rm dust}^{z=0}=36.5$ K and $\beta=1.5$ for Holmberg II, $T_{\rm dust}^{z=0}=23.5$ K and $\beta=1.5$ for DDO 165 \citep{Skibba11}. These are the galaxies shown in the left panel of Figure \ref{SED_IOK-1}. We adopted $T_{\rm dust}^{z=0}$'s  and $\beta$'s obtained by \citet{Skibba11} for all the galaxies shown in the right panel of Figure \ref{SED_IOK-1}. Then, to compute the factor $M_{\nu}$ at each frequency $\nu$ for each template galaxy, we scale the MBB with $T_{\rm dust}^{z=0}$ and $\beta$ to the peak of the FIR emission of the galaxy's SED and calculate the ratio $R_{\nu}$ of emission associated with the cold dust at each frequency $\nu$.
\begin{equation}
R_{\nu} = \frac{K \nu^{\beta} B_{\nu}(T_{\rm dust}^{z=0})}{F_{\nu}^{\rm int}}
\label{R_nu}
\end{equation}
where $K$ is the scaling factor. With this $R_{\nu}$, the equation (\ref{M_nu}) can be now expressed as 
\begin{equation}
M_{\nu} = (1 - R_{\nu}) + R_{\nu} \times \frac{B_{\nu}(T_{\rm dust}(z))}{B_{\nu}(T_{\rm dust}^{z=0})}
\label{M_nu2}
\end{equation}
Using the equations (\ref{Tdust}) and (\ref{C_nu})--(\ref{M_nu2}), we convert the original FIR continuum emission of each template galaxy at each frequency $\nu$ to the one including the effects of CMB and show the results in Figure \ref{SED_IOK-1}.

Comparison of the solid lines ({\it intrinsic} dust SEDs) and the dashed lines ({\it observed} dust SEDs which include the effects of CMB) in Figure \ref{SED_IOK-1} shows two notable differences. First, due to the extra heating of dust by CMB at $z=6.96$, the peaks of the SEDs shift toward shorter wavelengths (or higher frequencies). Also, the Rayleigh-Jeans tails of the dust emission observed against the CMB (dashed lines) are steeper than those of the intrinsically emitted SEDs (solid lines). This is because at a higher redshift CMB temperature is closer to the dust temperature, making the intrinsic dust emission more difficult to detect against CMB. The effects of CMB are stronger for galaxies with lower dust temperatures $T_{\rm dust}^{z=0}$. Especially, as the CMB temperature is $\sim 21.84$ K at $z \sim 7$, SEDs of the template galaxies with $T_{\rm dust}^{z=0}$ closer to this (i.e., M51, NGC 6946, NGC 0925, IC 2574 and DDO 165) are more remarkably affected by the CMB effects than those with $T_{\rm dust}^{z=0}$ much higher than 21.84 K. As discussed in \citet{daCunha13}, considering the effects of the CMB is required for the correct interpretation of dust continuum observations as it would change estimation of a dust continuum flux density and thus subsequent derivations of dust masses, total (F)IR luminosities and dust-obscured SFRs of galaxies.

Figure 2 also shows that the SED of IOK-1 is neither similar to those of highly dust-obscured galaxies such as Arp 220 and M82 \citep[see also][]{Walter12b,Gonzalez-Lopez14} nor less extreme dust-rich nearby spirals such as M51 and NGC 6946. It is rather more likely similar to SEDs of dwarfs/irregulars. More specifically, UV continuum slopes (see the insets in Figure \ref{SED_IOK-1}) and FIR continuum SEDs of IC 2574 (morphological type SABm), Holmberg II (Im), DDO 053 (Im), Holmberg I (IABm) and NGC 2915 (I0) are consistent with the detection data points of IOK-1 at rest frame UV and the ALMA FIR continuum upper limit \citep[morphological types were taken from][]{Dale07}. As local dwarf/irregular galaxies have been observed to have low IR luminosities and dust masses, IOK-1 could be such a system with low dust obscuration. Meanwhile, conducting SED-fitting to rest frame UV to optical fluxes, \citet{Ota10b} estimated the upper limit on the stellar mass of IOK-1 to be either $M_* \lesssim 2$--$9 \times 10^8 M_{\odot}$ for a young age ($\lesssim 10$ Myr) and low dust reddening ($A_V \sim 0$) or a mass as high as $M_* \lesssim 1$--$4 \times 10^{10} M_{\odot}$ for either an old age ($> 100$ Myr) or high dust reddening ($A_V \sim 1.5$). Given that LAEs at high redshifts (e.g., $z\sim5.7$--6.6) tend to be very young and low dust systems \citep[e.g.,][]{Ono10}, and considering the non-detection of IOK-1 in dust continuum with ALMA, the former $M_*$ limit would be more likely. For comparison, we have calculated the average stellar mass of 17 local dwarf and irregular galaxies (Sd and later) in \citet{Skibba11} KINGFISH survey sample and obtained $M_* \sim 9.1 \times 10^8 M_{\odot}$ (the stellar mass ranges from $2.2 \times 10^6 M_{\odot}$ to $5.8 \times 10^9 M_{\odot}$ within the sample). Hence, the stellar mass of IOK-1 is comparable to those of local dwarf/irregular galaxies. This also supports the idea that IOK-1 would be a system similar to local dwarf/irregular galaxies.

%%Table 2
%
%\begin{table}
%\begin{center}
%%\scriptsize
%\caption{Photometry of IOK-1\label{Photometry}}
%\begin{tabular}{cc}
%\tableline \tableline
%Waveband & Magnitude or Flux$^c$ \\
%\tableline
%$B$  & $>28.22$  \\
%$V$  & $>27.54$  \\
%$R$  & $>28.11$  \\
%$i'$ & $>27.35$  \\
%$z'_{\rm cont}$$^a$ & $27.63^{+1.7}_{-0.63}$ \\
%NB973 & $24.4\pm0.1$\\
%$y_{\rm cont}$$^a$  & $25.57^{+0.24}_{-0.20}$ \\
%F110W  & $25.50\pm0.04$ \\
%F125W  & $25.42\pm0.05$ \\
%F160W  & $25.44\pm0.05$ \\
%m$_{3.6\mu{\rm m}}$  &  \\
%m$_{4.5\mu{\rm m}}$  &  \\
%m$_{5.8\mu{\rm m}}$  & $>21.83$ \\
%m$_{8.0\mu{\rm m}}$  & $>21.52$ \\
%$S_{250\mu{\rm m}}$  & $<13.8$ \\
%$S_{350\mu{\rm m}}$  & $<15.9$ \\
%$S_{500\mu{\rm m}}$  & $<18.0$ \\
%$S_{1.26{\rm mm}}$$^b$ &  \\
%$S_{\rm [CII]}$$^b$ &  \\
%\tableline
%\end{tabular}
%\tablenotetext{a}{UV continuum magnitudes calculated by subtracting the Ly$\alpha$ line flux measured in the spectrum of IOK-1 from $z'$ and $y'$ band total fluxes (see Ota et al.~2010b for details).}
%\tablenotetext{b}{They correspond to $3\sigma$ FIR continuum and [CII] line sensitivities of our ALMA observation.}
%\tablenotetext{c}{The unit of flux is total AB magnitude for $B$ to 8.0$\mu$m and mJy for 24$\mu$m to $f_{\rm [CII]}$. All the limits are $3\sigma$. The fluxes in $B$ to $y_{\rm cont}$ as well as 5.8$\mu$m and 8.0$\mu$m were taken from Ota et al.~(2010b), F110W to F160W from Jiang et al.~(2013) and 3.6$\mu$m and 4.5$\mu$m from Egami et al.~(2013).}  
%\end{center}
%\end{table}

\subsubsection{Dust Mass, IR Luminosity and Dust-obscured SFR \label{FIR_Properties}}
Given our $3\sigma$ ALMA upper limit on 1.26 mm continuum flux density $F_{\nu/(1+z)}$, we can estimate the upper limit on the dust mass $M_{\rm dust}$ of IOK-1 from the following relation. 
\begin{eqnarray}
M_{\rm dust} &=& \frac{F_{\nu/(1+z)} D_{\rm L}^2}{(1+z)\kappa_{\rm d}(\nu_{\rm rest})B_{\nu}(T_{\rm dust})} \nonumber \\
             &=& \frac{F_{\nu/(1+z)}^{\rm int} D_{\rm L}^2}{(1+z)\kappa_{\rm d}(\nu_{\rm rest})B_{\nu}(T_{\rm dust}^{z=0})} \nonumber \\ 
             &=& \frac{F_{\nu/(1+z)}^{\rm obs} D_{\rm L}^2}{(1+z)\kappa_{\rm d}(\nu_{\rm rest})C_{\nu}M_{\nu}B_{\nu}(T_{\rm dust}^{z=0})} \nonumber \\
             &=& \frac{F_{\nu/(1+z)}^{\rm obs} D_{\rm L}^2}{(1+z)\kappa_{\rm d}(\nu_{\rm rest})[B_{\nu}(T_{\rm dust}(z)) - B_{\nu}(T_{\rm CMB}(z))]} \nonumber \\
\label{DustMass}
\end{eqnarray}
Here, $D_{\rm L}$ is a luminosity distance, $\kappa_{\rm d}(\nu_{\rm rest})$ is the rest frame frequency dust mass absorption coefficient, and $B_{\nu}(T_{\rm dust})$ is the Planck function at a rest frame frequency $\nu_{\rm rest}$ and a dust tempatature $T_{\rm dust}$. We adopt $\kappa_{\rm d}(\nu_{\rm rest})=0.77 (850\mu{\rm m}/\lambda_{\rm rest})^{\beta}$ cm$^2$ g$^{-1}$ from Dunne et al.~(2000) with the rest frame wavelength of $\lambda_{\rm rest}=158$ $\mu$m. $F_{\nu/(1+z)}^{\rm int}$ is the intrinsic continuum flux that can be obtained after correcting the observed continuum flux $F_{\nu/(1+z)}^{\rm obs} = 63$ $\mu$Jy for the CMB effects by using the equation (\ref{F_int}). Because the SED of IOK-1 is similar to those of local dwarf and irregular galaxies (see \textsection \ref{SED} and Figure \ref{SED_IOK-1}), we model the dust continuum SED of IOK-1 as an MBB with $T_{\rm dust}^{z=0}=27.6$ K and $\beta=1.5$, the average values of 17 local dwarf and irregular galaxies (Sd and later) derived by \citet{Skibba11} from their KINGFISH survey. Then, the equation (\ref{DustMass}) gives $M_{\rm dust} < 6.4 \times 10^{7} M_{\odot}$. The upper limit on dust mass estimated here is also listed in Table 1. If we do not correct the effects of CMB (i.e., if we use the observed continuum flux limit $F_{\nu/(1+z)}^{\rm obs} = 63$ $\mu$Jy with $B_{\nu}(T_{\rm dust}^{z=0})$ instead of $C_{\nu}M_{\nu}B_{\nu}(T_{\rm dust}^{z=0})$ for the equation (\ref{DustMass})), this leads to $M_{\rm dust}^{\rm CMB} < 4.9 \times 10^{7} M_{\odot}$, a factor of 0.76 ($=C_{\nu}M_{\nu}$) underestimation compared to the intrinsic dust mass limit. Meanwhile, if  $T_{\rm dust}^{z=0}$ of IOK-1 was actually lower (higher) than what we assumed here (27.6 K), the upper limit on $M_{\rm dust}$ would be higher (lower).
%For comparison, if we let $F_{\nu/(1+z)} = F_{\nu/(1+z)}^{\rm obs} = XX$ $\mu$Jy obtained by using the template of NGC 6946 as an approximation of the dust continuum SED, then we obtain $M_{\rm dust}^{\rm int} < XX M_{\odot}$.  

We also estimate the upper limit on intrinsic FIR luminosity $L_{\rm FIR}$ of IOK-1 by again modeling its dust continuum SED as an MBB with $T_{\rm dust}^{z=0}=27.6$ K and $\beta=1.5$ \citep[the average values of local dwarfs and irregulars derived by][see above]{Skibba11}, scaling it to our CMB-corrected $3\sigma$ ALMA upper limit on 1.26 mm continuum (i.e., $F_{\nu/(1+z)}^{\rm obs}/C_{\nu}M_{\nu} = F_{\nu/(1+z)}^{\rm int}$) and integrating it from 42.5 $\mu$m to 122.5 $\mu$m. We obtain $L_{\rm FIR} < 3.7 \times 10^{10} L_{\odot}$. Moreover, integrating the same scaled MBB from 8 $\mu$m to 1 mm, we estimate the intrinsic total IR luminosity to be $L_{\rm IR} < 5.7 \times 10^{10} L_{\odot}$. These intrinsic luminosities correspond to only the contribution from the stars in IOK-1 to dust heating and do not include the contribution from CMB. Hence, especially, $L_{\rm IR}$ is appropriate for estimating the dust-obscured SFR. If we use the \citet{Kennicutt98} relation with the \citet{Salpeter55} initial mass function (IMF), SFR$_{\rm dust}= 1.73 \times 10^{-10} L_{\rm IR}$ $[L_{\odot}]$, the total IR luminosity converts to the dust-obscured SFR of SFR$_{\rm dust} < 10.0 $ $M_{\odot}$yr$^{-1}$. \citet{Jiang13} estimated the SFR of IOK-1 to be SFR$_{\rm UV} \sim 23.9$ $M_{\odot}$yr$^{-1}$ from the HST/WFC3 band magnitudes which cover the rest frame UV continuum of IOK-1 not including Ly$\alpha$ emission. This is the SFR not obscured by dust. Hence, the upper limit on the total (obscured plus un-obscured) SFR of IOK-1 is SFR$_{\rm total} =$ SFR$_{\rm UV} +$ SFR$_{\rm dust} < 33.9$ $M_{\odot}$yr$^{-1}$. This implies that $f_{\rm obscured}=$ SFR$_{\rm dust}/$SFR$_{\rm total} < 29$\% of star formation is obscured by dust in IOK-1. On the other hand, if we scale the MBB to just the observed continuum flux density limit $F_{\nu/(1+z)}^{\rm obs}$ (i.e., the flux not corrected for the CMB effects) and integrate it, this would result in underestimations of $L_{\rm FIR}$, $L_{\rm IR}$ and SFR$_{\rm dust}$ by 24\% (a factor of $C_{\nu}M_{\nu}=0.76$) compared to the intrinsic ones we have obtained above, due to the extra heating of dust by CMB and the reduction of observed dust continuum flux by the CMB as a background.

The dust mass and FIR/IR luminosity of IOK-1 are $\sim 1$--2 orders of magnitude lower than those of a host galaxy of the quasar ULAS J112001.48+064124.3 at a similar redshift $z=7.08$ \citep{Venemans12} and the most distant SMG known, HFLS 3, at $z=6.34$ \citep{Riechers13} as well as those of the quasar host galaxies at $z\sim1.8$--6.4 studied so far \citep{Pety04,Maiolino05,Beelen06,Iono06,Maiolino09,Wagg10,Wang13,Willott13} and $z\sim2$--5 SMGs \citep{Riechers10,Dwek11,Cox11,DeBreuck11,Valtchanov11,Walter12a}. Furthermore, though the sensitivities are about one order of magnitude lower than that of ALMA, \citet{Walter12b} also observed IOK-1 in dust continuum, \citet{Kanekar13} and \citet{Gonzalez-Lopez14} carried out 1.2 mm continuum observations of three other $z\sim6.6$ LAEs, and they all ended up with non-detections. Our and their results together imply that high redshift LAEs at the epoch of reionization are likely systems with very little dust, quite different in dust content from extreme starbursts or highly dust-obscured galaxies such as quasar hosts and SMGs at similar and lower redshifts. 

%%figure 4

\begin{figure}[t]
\epsscale{1.17}
\plotone{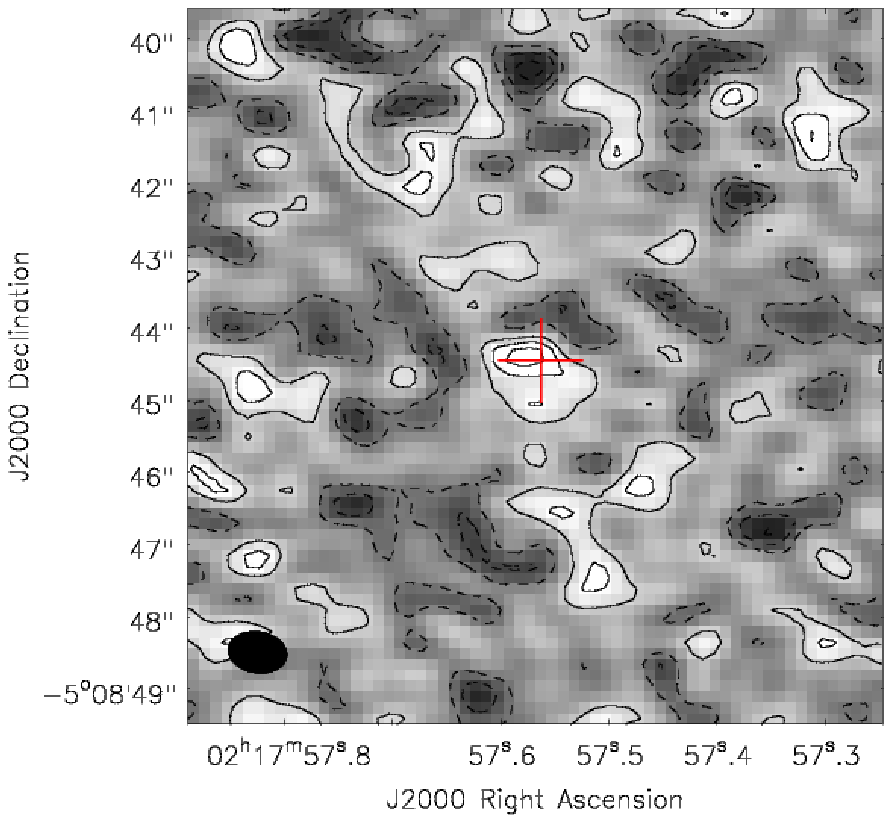}
\caption{The rest frame 158 $\mu$m FIR continuum image of Himiko we re-reduced at a resolution of $0\farcs8 \times 0\farcs5$. Coordinates are in the J2000.0 system and contours are in steps of $1\sigma =$ 19.7 $\mu$Jy beam$^{-1}$, starting at $\pm 1\sigma$. Positive and negative contours are presented with solid and dashed lines, respectively. The synthesized beam is shown in the bottom left. The position of Himiko measured in the optical narrowband NB921 (Ly$\alpha$ line plus rest frame UV continuum) image \citep{Ouchi09a} is marked with the cross. There is a $3.4\sigma$ possible weak continuum source at the position of Himiko with a peak flux of 67.5 $\mu$Jy. \label{HIMIKO}}
%Note that due to our tunig the systemic redshift corresponds to a velocity of $-150$ km s$^{-1}$ in this representation (See Section 2).
\end{figure}

%%figure 5

\begin{figure*}[t]
%\begin{figure*}
\epsscale{1.15}
%\plotone{f5.eps}
\plotone{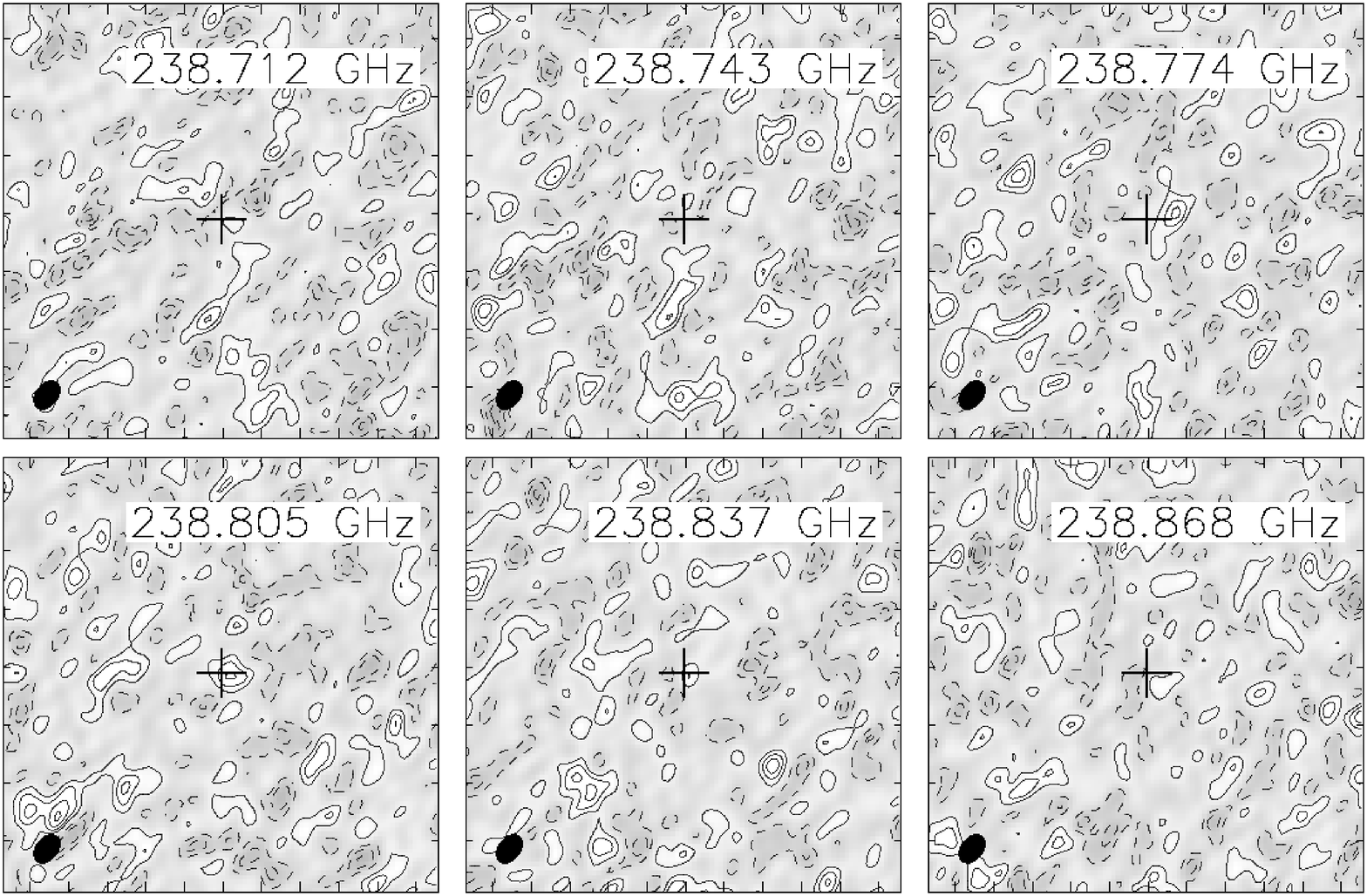}
\caption{The [CII] channel maps of IOK-1 at a resolution of $1\farcs1 \times 0\farcs75$ around the expected redshift. Contours are shown in steps of $1\sigma = 215$ $\mu$Jy beam$^{-1}$ starting at $\pm1\sigma$. The central frequency is shown in each panel. Positive and negative contours are presented with solid and dashed lines, respectively. The synthesized beam is shown in the bottom left. The position of IOK-1 measured in the optical narrowband NB973 (Ly$\alpha$ line plus rest frame UV continuum) image \citep{Iye06,Ota08} is marked with the cross. A tentative $3\sigma$ signal is seen in the bottom left panel.\label{CII_IOK1image}}
%Note that due to our tunig the systemic redshift corresponds to a velocity of $-150$ km s$^{-1}$ in this representation (See Section 2).
\end{figure*}

Moreover, \citet{Ouchi13} observed dust continuum of a $z=6.595$ LAE, Himiko, which has a similar SFR$_{\rm UV}$ ($\sim 30$ $M_{\odot}$yr$^{-1}$) to IOK-1, with ALMA to a similar sensitivity to ours at $0\farcs82 \times 0\farcs58$ resolution but found only a $\sim 2\sigma$ weak signal at the expected position and considered it a non-detection. Thus, Himiko could be similar to typical high redshift LAEs in dust content. However, Himiko has at least twice larger size than IOK-1 in Ly$\alpha$ ($\gtrsim 3''$ in extent) and is considered a Ly$\alpha$ blob (LAB). Also, its UV continuum shows a few knots over about $2''$ \citep{Ouchi13}. Hence, to search for possible extended emission, we have re-calibrated, flagged, and imaged the public Himiko ALMA Cycle 0 data. We made the continuum map that is based on all data except the SPW in which the [CII] line is expected, to avoid a possible contamination from that line. As seen in Figure \ref{HIMIKO}, at a nearly full resolution of $0\farcs8 \times 0\farcs5$, we found that there is a $3.4\sigma$ possible weak continuum source at the position of Himiko, as pointed out by \citet{Ouchi13}. The source shows a possible extension East-West, along the UV continuum direction, although low signal to noise ratio does not allow us to confirm if the source is extended. If we consider this $3.4\sigma$ signal the upper limit (as it is marginal) and model the dust SED of Himiko as an MBB with $T_{\rm dust}^{z=0}=27.6$ K and $\beta=1.5$ typical of dwarf/irregular galaxies, integrations of the MBB scaled to 67.5 $\mu$Jy/$C_{\nu}M_{\nu}$ ($C_{\nu}M_{\nu}\sim 0.80$ for Himiko) give $L_{\rm FIR} < 3.5 \times 10^{10} L_{\odot}$ and $L_{\rm IR} < 5.4 \times 10^{10} L_{\odot}$. The \citet{Kennicutt98} relation converts the $L_{\rm IR}$ to SFR$_{\rm dust} < 9.3$ $M_{\odot}$yr$^{-1}$ and SFR$_{\rm total} < 39.3$ $M_{\odot}$yr$^{-1}$, implying that $f_{\rm obscured} < 24$\% of star formation is obscured by dust. Also, the equation (\ref{DustMass}) gives dust mass of Himiko, $M_{\rm dust} <  6.1 \times 10^7$ $M_{\odot}$.
%While Himiko has an SFR$_{\rm UV} \sim 30 M_{\odot}$ yr$^{-1}$, similar order as the SFR$_{\rm UV}$ of IOK-1, SED-fitting analysis at rest frame UV to optical wavelengths implies a higher total SFR, SFR$_{\rm total} \sim 100 M_{\odot} {\rm yr}^{-1}$ (Ouchi et al.~2013). Also, Himiko is considered a Ly$\alpha$ blob and has at least twice larger size than IOK-1 in Ly$\alpha$. 
%Himiko is significantly extended in Ly$\alpha$ and UV continuum emission. The Ly$\alpha$ appears to be at least $3''$ in extent, while the UV continuum shows a few knots over about $2''$. The dust continuum observation with ALMA had a resolution of $0\farcs82 \times 0\farcs58$ (Ouchi et al.~2013). 
%At a nearly full resolution of $0\farcs8 \times 0\farcs5$, we obtain the rms of 19.7 $\mu$Jy beam$^{-1}$. We show this continuum map in Figure \ref{HIMIKO}, and as pointed out by Ouchi et al.~(2013), we found that there is a $3.4\sigma$ possible weak continuum source at the position of Himiko with a peak flux of 67.5 $\mu$Jy. 
%However, note that as the $3.4\sigma$ signal at the position of Himiko is marginal, a deeper observation is necessary to definitvely tell if it is indeed a detection of the dust continuum.  

Meanwhile, imaging at $2\farcs01 \times 1\farcs85$ resolution (comparable to the Ly$\alpha$ size of Himiko) resulted in a higher noise, and a non-detection with an rms of 42 $\mu$Jy beam$^{-1}$. Integrations of the same MBB scaled to the $3\sigma$ limit give the conservative upper limits, $L_{\rm FIR} < 6.5 \times 10^{10} L_{\odot}$, $L_{\rm IR} < 1.0 \times 10^{11} L_{\odot}$, SFR$_{\rm dust} < 17.4$ $M_{\odot}$yr$^{-1}$, SFR$_{\rm total} < 47.4$ $M_{\odot}$yr$^{-1}$ and $f_{\rm obscured}< 37$\%. Also, the equation (\ref{DustMass}) gives the conservative dust mass upper limit, $M_{\rm dust} < 1.1 \times 10^8$ $M_{\odot}$. Hence, although it is not clear that Himiko is extended in dust continuum, it could be similar to other common LAEs at high redshifts such as IOK-1 in dust content or possibly slightly dustier if the $3.4\sigma$ signal is a real detection. It should be noted that comparison of these continuum flux limits at $0\farcs8 \times 0\farcs5$ and $2\farcs01 \times 1\farcs85$ resolutions with dust SEDs of different types of local galaxies in Figure 5 in \citet{Ouchi13} implies that Himiko could be a system similar to local dwarf/irregular galaxies, not changing the same implication obtained by \citet{Ouchi13} by using their derived Himiko continuum limit. Thus, our assumption of the MBB with $T_{\rm dust}^{z=0}$ and $\beta$ typical of local dwarf/irregular galaxies for Himiko to derive its dust-related properties is reasonable.

\begin{figure}
\epsscale{1.2}
%\plotone{CIIspec.eps}
%\plotone{fig5.eps}
\plotone{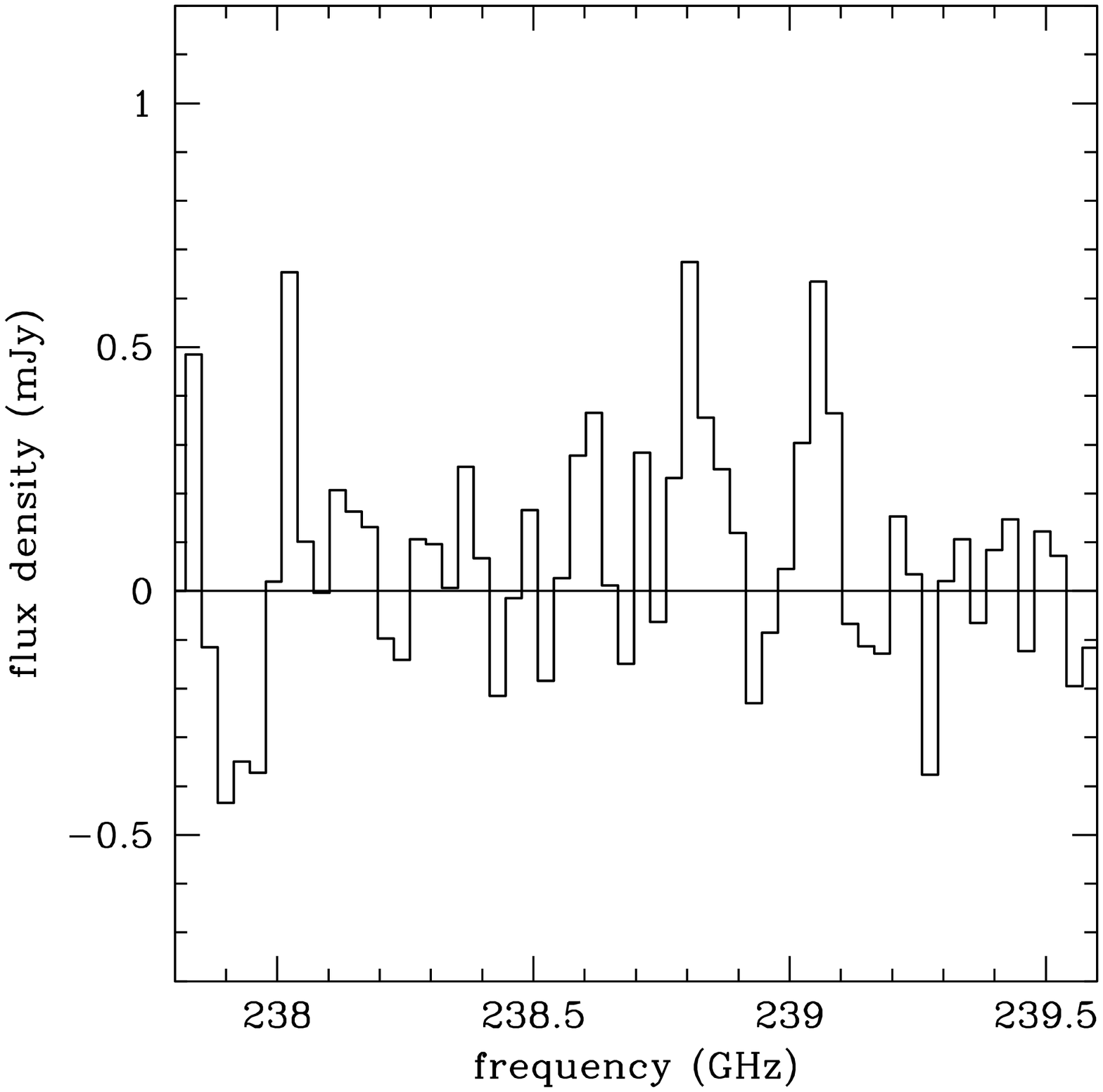}
\caption{The spectrum at the position of IOK-1 at a spatial resolution of $1\farcs1 \times 0\farcs75$ and a spectral resolution of 40 km s$^{-1}$. There is a $3\sigma$ weak signal at $\sim 238.8$ GHz, and this corresponds to $z=6.9587$ if this was the [CII] line. It would be slightly lower than the redshift measured from Ly$\alpha$ emission peak of IOK-1: $z_{{\rm Ly}\alpha}=6.96$ \citep{Iye06}; $z_{{\rm Ly}\alpha}=6.965$ \citep{Ono12}. This would be consistent with the idea that the $z_{{\rm Ly}\alpha}$ could be overestimated due to the absorption of blue side of Ly$\alpha$ emission by the intergalactic medium. However, the $3\sigma$ signal is marginal and not strong enough to definitively conclude that it is indeed the [CII] line. Hence, we treat it as non-detection in this paper. There is a similar feature at 239.05 GHz. However, this portion of the spectrum has higher noise than the one at $< 239.0$ GHz due to atmospheric absorption.\label{CII_IOK1spec}}
\end{figure}

\subsection{[CII] Line Properties}
Figures \ref{CII_IOK1image} and \ref{CII_IOK1spec} show the [CII] channel maps and the spectrum at a resolution of $1\farcs1 \times 0\farcs75$ around the expected redshift of IOK-1 based on Ly$\alpha$ emission. We see a $3 \sigma$ weak signal at the expected position of IOK-1 with a peak at $\sim 238.805$ GHz (the peak flux density is $\sim 670$ $\mu$Jy) and an FWHM of $\sim 0.05$ GHz or $\sim 63$ km s$^{-1}$ (there are also similar $3\sigma$ features at 239.05 GHz and 238.02 GHz, and we will mention these below). If the signal at $\sim 238.805$ GHz was indeed the [CII] line, the redshift based on its peak is $z_{\rm [CII]}=6.9585$. It is slightly lower than the redshift measured from Ly$\alpha$ emission peak: $z_{{\rm Ly}\alpha}=6.96$ \citep[][using Subaru Telescope FOCAS spectrograph]{Iye06}; $z_{{\rm Ly}\alpha}=6.965$ \citep[][using Keck Telescope DEIMOS spectrograph with a slightly better spectral resolution]{Ono12} by $\sim 57$--245 km s$^{-1}$. This is consistent with the idea that $z_{{\rm Ly}\alpha}$ could be overestimated due to the absorption of blue side of Ly$\alpha$ emission by the intergalactic medium. Also, previous spectroscopic studies of LAEs and LBGs at $z\sim2$--3 have shown that their Ly$\alpha$ emission peaks are usually shifted redward of the redshifts accurately measured from other lines and absorptions \citep[e.g.,][]{Pettini01,Shapley03,Steidel10,McLinden11,Shibuya14}. IOK-1 is not only an LAE detected in a narrowband \citep{Iye06,Ota08} but also an LBG detected as a $z'$-band dropout \citep{Ouchi09b,Ono12}. Thus, if the $3\sigma$ feature is indeed the [CII] line, it is also consistent with the trend that redshifts of LAEs and LBGs measured from Ly$\alpha$ tend to be higher than their systemic redshifts. However, the $3\sigma$ significance of this signal is marginal and is not strong enough to definitively conclude that the detected feature is indeed the [CII] line. Hence, in this paper, we consider it a non-detection and leave the firm conclusion about detection/non-detection to our forthcoming (scheduled) observations of IOK-1 with ALMA. Note that there are also similar $3\sigma$ features at 239.05 GHz and 238.02 GHz. However, as for the one at 239.05 GHz, this portion of the spectrum is noisier than the one at frequencies $< 239.0$ GHz due to atmospheric absorption. The feature at 238.02 GHz is unlikely to be [CII] line, because if it was [CII], the redshift derived from it would be higher than the redshift estimated from Ly$\alpha$ emission.

As we mentioned earlier, IOK-1 has a size of $\sim 1\farcs6 \times 1\farcs0$ in Ly$\alpha$, larger than the beamsize of the [CII] map $1\farcs1 \times 0\farcs75$. Hence, we also made the [CII] map convolved to a $1\farcs5 \times 1\farcs2$ resolution (through tapering with a 100 k$\lambda$ taper) comparable to the Ly$\alpha$ size to see if we detect any extended source. However, we did not detect IOK-1 in this map, either. As the rms of this map is $\sigma_{\rm cont}=240$ $\mu$Jy beam$^{-1}$, the $3\sigma$ upper limit on [CII] line flux density of IOK-1 is 720 $\mu$Jy beam$^{-1}$ over a channel width of 40 km s$^{-1}$ (same as the highest spectral resolution at our observing frequency of 238.76 GHz in the TDM mode we used). This converts to a line luminosity of $L_{\rm [CII]} < 3.4 \times 10^7 L_{\odot}$. We assumed a line width of $\Delta v = 40$ km s$^{-1}$. \citet{Gonzalez-Lopez14} simulated the [CII] line of IOK-1 by using the same procedure as the one of \citet{Vallini13} who combined high resolution, radiative transfer cosmological simulations of $z\sim6$ galaxies with a sub-grid multi-phase model of their interstellar medium and derived the expected intensity of several FIR emission lines including [CII] for different values of the gas metallicity. Conservatively assuming solar metallicity, their simulation suggests that most of the [CII] emission of IOK-1 comes from a cold neutral medium and that FWHM of its main peak is $\sim 50$ km s$^{-1}$. In this context, our assumption of $\Delta v = 40$ km s$^{-1}$ for calculating the [CII] line luminosity upper limit is reasonable. In this section, based on this limit, we explore the properties of IOK-1 related to the [CII] line. Note that the effects of CMB on [CII] line observations of high redshift galaxies were discussed and found to be negligible by \citet{Gonzalez-Lopez14}. Hence, we do not consider the effects of CMB on [CII] line in the following.   
%Hence, there is a possibility that the $3\sigma$ signal might be the detection of [CII] line. However, 

%%figure 7

\begin{figure*}
%\epsscale{1.22}
%\plotone{Figure_LCII_SFR_V3_50kmps_40kmps.eps}
%\plotone{Figure_LCII_SFR_V3_50kmps_40kmps-HimikoSEDfitLegend.eps}
%\plotone{Figure_LCII_SFR_V3_50kmps_40kmps-HimikoSEDfitLegend-2beamsIOK1-11beamsHimiko.eps}
%\plotone{Figure_LCII_SFR_V3_50kmps_40kmps-HimikoSEDfitLegend-taperedIOK1-taperedHimiko.eps}
\plotone{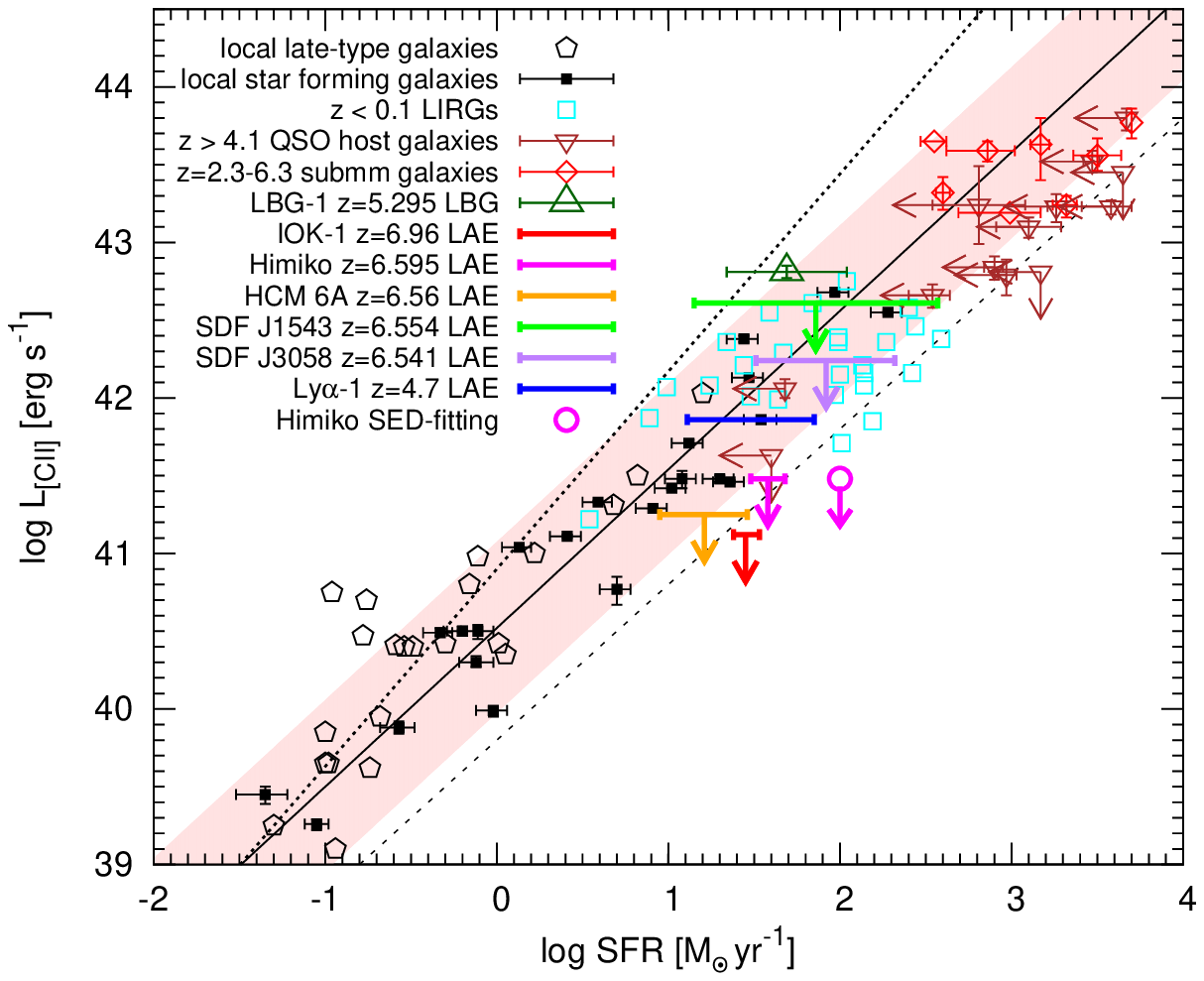}
\caption{[CII] luminosity $L_{\rm [CII]}$ as a function of SFR compiled for different types of objects (arrows indicate upper limits): local late-type galaxies with their correlation (dotted line) between total SFR and $L_{\rm [CII]}$ \citep{Boselli02}, local star-forming galaxies with their correlation (solid line with shaded region) between total SFR and $L_{\rm [CII]}$ with $2\sigma$ scatter \citep{DeLooze11}, $z<0.1$ LIRGs \citep{Maiolino09}, $z=4.1$--7.1 quasar host galaxies \citep{Pety04,Maiolino05,Iono06,Maiolino09,Wagg10,Venemans12,Wang13,Willott13}, $z=2.3$--6.34 submm galaxies \citep{Iono06,Ivison10,Cox11,DeBreuck11,Valtchanov11,Walter12a,Riechers13,Rawle14} and a $z=5.295$ triple LBG system named LBG-1 associated with a $z\sim 5.3$ protocluster \citep{Capak11,Riechers14}. SFRs of LIRGs, quasar host galaxies and submm galaxies are those converted from their total FIR or IR luminosities (depending on availability in the references) by either us or authors of the references using the \citet{Kennicutt98} relation. SFRs of quasar host galaxies are upper limits as their total F(IR) luminosities might include AGN contribution. The dashed line is the SFR--$L_{\rm [CII]}$ relation derived for $L_{\rm FIR} > 10^{12} L_{\odot}$ sources by \citet{Maiolino05}. We also plot the possible range of total SFRs (see text in \textsection \ref{CII_SFR}) and $L_{\rm [CII]}$ (limits) of $z\simeq 4.7$--7 LAEs with the color coded lines and arrows: IOK-1 (this study with a velocity width of $\Delta v= 40$ km s$^{-1}$), Himiko \citep[the data taken by][and re-reduced by us; We plot the $L_{\rm [CII]}$ limit at $2.''5$ resolution and $\Delta v= 36$ km s$^{-1}$, and SFR$_{\rm dust}$ at $2.''01 \times 1.''85$ resolution; See text in \textsection \ref{CII_SFR}]{Ouchi13}, HCM 6A \citep[][$\Delta v= 100$ km s$^{-1}$]{Kanekar13}, SDF J132408.3+271543, SDF J132415.7+273058 \citep[][$\Delta v= 50$ km s$^{-1}$]{Gonzalez-Lopez14} and Ly$\alpha$-1 \citep{Carilli13}. The open circle shows the Himiko $L_{\rm [CII]}$ limit we derived with its total SFR estimated by the SED-fitting \citet{Ouchi13} conducted. All the SFRs plotted here are based on Salpeter IMF. If the SFRs in literature are those based on \citet{Chabrier03} or \citet{Kroupa01} IMFs, we multiplied them by factors 1.6 and 1.5, respectively \citep[e.g.,][]{Elbaz07,Salim07}, to convert them to SFRs with Salpeter IMF. \label{SFR_vs_LCII}}
\end{figure*}

\subsubsection{[CII] Line and SFRs of High Redshift LAEs\label{CII_SFR}}
The [CII] line has been considered a tracer of SFR as it is produced by the UV field in star forming regions. Figure \ref{SFR_vs_LCII} shows the [CII] luminosity as a function of SFR for several different types of objects. Objects with higher SFRs have higher [CII] luminosities, and different types of objects reside in different locations in the SFR--$L_{\rm [CII]}$ plane in such a way that local late-type and star-forming galaxies are located at the low to intermediate SFR and $L_{\rm [CII]}$ regime, low redshift LIRGs at the intermediate to high SFR and $L_{\rm [CII]}$ region and quasar host galaxies and SMGs at the very high SFR and $L_{\rm [CII]}$ domain. For the local late-type and star-forming galaxies, clear correlations have been observed to hold between SFR and $L_{\rm [CII]}$ \citep{Boselli02,DeLooze11}. In Figure \ref{SFR_vs_LCII}, we compare these calibrated relations for local galaxies and locations of all types of objects in the SFR--$L_{\rm [CII]}$ plane with total SFRs and [CII] luminosities of IOK-1 as well as other LAEs at $z\sim4.7$--6.6 previously observed to see whether they are similar or different systems in gas enrichment or in metallicity as [CII] luminosity would be proportional to metallicity \citep[e.g.,][]{Vallini13}.

In \textsection \ref{FIR_Properties}, we have estimated an upper limit on the total SFR of IOK-1 to be SFR$_{\rm total} < 33.8$ $M_{\odot}$yr$^{-1}$ by adding the SFR estimated from its UV continuum (SFR$_{\rm UV}$) and the upper limit on the dust-obscured SFR estimated from 1.26 mm continuum (SFR$_{\rm dust}$). If IOK-1 had no dust, the lowest possible total SFR would be the one estimated from its UV continuum, i.e., SFR$_{\rm total} \geq$ SFR$_{\rm UV} = 23.9$ $M_{\odot}$yr$^{-1}$ \citep{Jiang13}. In Figure \ref{SFR_vs_LCII}, we also plot the possible total SFR range of IOK-1 and the upper limit on [CII] luminosity. We notice that IOK-1 neither seems to follow the SFR--$L_{\rm [CII]}$ correlations for local star-forming galaxies (though the correlation has some scatters) nor to be within the regions occupied by any other type of objects except for high redshift LAEs (i.e., Himiko and HCM 6A which we will mention below). Its [CII] luminosity is significantly fainter than that expected for a local star-forming galaxy having the same SFR as IOK-1. By using the local star-forming galaxy correlation derived by \citet{DeLooze11}, the [CII] luminosity limit of IOK-1 converts to SFR of $\sim 2.5$ $M_{\odot}$yr$^{-1}$ (or $\log$ SFR $\sim 0.51$), which is much lower than  the possible SFR range of IOK-1, 23.9--33.8 $M_{\odot}$yr$^{-1}$. Hence, the local star-forming galaxy correlation does not seem to apply to IOK-1. IOK-1 could be a system having either lower enriched gas or different photodissociation region (PDR) structure/physical state of ISM than local star-forming galaxies.

A similar result has been reported for a $z=6.595$ LAE, Himiko, by \citet{Ouchi13} who did not detect its [CII] line by using ALMA at $0\farcs82 \times 0\farcs58$ resolution and 200 km s$^{-1}$ channel$^{-1}$. As Himiko is extended over $\gtrsim 3''$ in Ly$\alpha$, we generated the spectrum at $2.''5$ resolution and 30 MHz channel$^{-1}$ (36 km s$^{-1}$), using the public Himiko ALMA Cycle 0 data, to see if we detect any extended source. As with the higher resolution analysis of \citet{Ouchi13}, no line is detected. The 3$\sigma$ flux density upper limit at $2.''5$ resolution is 2.0 mJy beam$^{-1}$ channel$^{-1}$ at 36 km s$^{-1}$ channel$^{-1}$ or $L_{\rm [CII]} < 7.9 \times 10^7 L_{\odot}$ (assuming a velocity width of $\Delta v =36$ km s$^{-1}$), and a factor 1.7 lower at 100 km s$^{-1}$ channel$^{-1}$. However, we have found that the rms noise is about a factor two higher than this between 249.7 GHz to 250.2 GHz due to atmospheric absorption, so if the line happened to fall in this frequency range, the limits are a factor of two worse. In \textsection \ref{FIR_Properties}, we also derived the conservative upper limit on SFR$_{\rm total}$ of Himiko at $2\farcs01 \times 1\farcs85$ resolution (comparable to the Ly$\alpha$ size of Himiko). In Figure \ref{SFR_vs_LCII}, we plot the $L_{\rm [CII]}$ upper limit and the possible SFR range of Himiko (SFR$_{\rm UV} \sim 30$ $M_{\odot}$yr$^{-1} \leq$ SFR$_{\rm total} < $ SFR$_{\rm UV}$ $+$ SFR$_{\rm dust} \sim 47.4$ $M_{\odot}$yr$^{-1}$). Himiko is off the local galaxy SFR--$L_{\rm [CII]}$ correlation and close to IOK-1. For comparison, in Figure \ref{SFR_vs_LCII}, we also plot the $L_{\rm [CII]}$ upper limit of Himiko we derived with SFR$_{\rm total}=100$ $M_{\odot}$yr$^{-1}$ which \citet{Ouchi13} estimated from SED-fitting at rest frame UV to optical wavelengths. In this case, Himiko is also off the local galaxy SFR--$L_{\rm [CII]}$ correlation but apart from IOK-1 due to the difference in SFR$_{\rm total}$.

Both IOK-1 and Himiko have several things in common except that Himiko displays more extended Ly$\alpha$ emission and is considered an LAB and that Himiko might be detected in dust continuum. They both consist of 2--3 merging components \citep{Cai11,Ouchi13}, have similar total UV continuum luminosities (or SFR$_{\rm UV}$) of the whole system and have not been detected in [CII] line to the similar flux limits (if we assume that the sizes of these LAEs in [CII] are similar). Hence, if SFR$_{\rm UV}$ $+$ SFR$_{\rm dust}$ more likely reflects the actual total SFR of Himiko than the SFR estimated from the SED-fitting, both IOK-1 and Himiko may be normally star-forming galaxies at high redshifts with lower enriched gas/metallicity content than local normal star-forming galaxies. 

On the other hand, \citet{Kanekar13} observed a gravitationally lensed $z=6.56$ faint LAE, HCM 6A, in [CII] emission with the IRAM/PdBI to a deep flux limit but somewhat shallower than ours and Ouchi et al.'s taking advantage of the lensing magnification. They could not detect [CII] line, either. Given SFR$_{\rm UV} \sim 9$ $M_{\odot}$yr$^{-1}$ \citep{Hu02a,Hu02b}, SFR$_{\rm dust} < 19.6$ $M_{\odot}$yr$^{-1}$ (we derived this SFR in the same way as we did for IOK-1 and Himiko for a fair comparison) and the $L_{\rm [CII]}$ upper limit \citep[they adopted 100 km s$^{-1}$ width]{Kanekar13}, we also plot HCM 6A in Figure \ref{SFR_vs_LCII}. It is close to the local SFR--$L_{\rm [CII]}$ relation although it could be off the relation as the [CII] luminosity is an upper limit. This is possibly due to their slightly shallower [CII] flux limit. Since this LAE is a factor of 2--3 lower than IOK-1 and Himiko in SFR$_{\rm UV}$, its [CII] luminosity could also potentially be lower than those of IOK-1 and Himiko.
%, if [CII] line is also a good tracer of SFR for high redshift LAEs just like it is for local galaxies. If this is the case, a deeper observation of HCM 6A might reveal that it would be actually apart from the local relation. If HCM 6A is below the local relation, since this LAE is among the faintest and IOK-1 and Himiko are among the brightest (in Ly$\alpha$ emission) LAEs at $z\sim7$ and $z\sim6.6$ detected with Ly$\alpha$ narrowband filters to date, commom high redshift LAEs detected similarly with narrowbands (i.e., LAEs brighter than HCM 6A and fainter than IOK-1 and Himiko) could be also a system with lower enriched gas/metallicity content than local star-forming galaxies. However, we definitely need deeper surveys with more statistics to confirm this.

In Figure \ref{SFR_vs_LCII}, we also plot the $z=6.554$ and $z=6.541$ LAEs, SDF J132408.3+271543 and SDF J132415.7+273058 \citep{Taniguchi05}, whose [CII] lines and 1.2 mm continua were observed by \citet{Gonzalez-Lopez14} with CARMA and undetected. We plot the SFR ranges estimated from SFR$_{\rm UV}$'s and the SFR$_{\rm dust}$ upper limits (corrected for the CMB effects on their 1.2 mm continua) derived by \citet{Jiang13} and \citet{Gonzalez-Lopez14}, respectively. They are within the local SFR-$L_{\rm [CII]}$ relation and the low redshift LIRG region, but the $L_{\rm [CII]}$ upper limits (50 km s$^{-1}$ channel width) obtained by \citet{Gonzalez-Lopez14} are an order of magnitude shallower than those for IOK-1, Himiko and HCM 6A. Therefore, at this moment, we can neither tell whether these two LAEs follow the local relation nor whether these results are consistent with those of IOK-1, Himiko and HCM 6A. Deeper observations are required to reveal this. 

On the contrary, a $z\sim4.7$ LAE, Ly$\alpha$-1, observed with ALMA, is detected in [CII] line \citep{Carilli13} and located in the local SFR--$L_{\rm [CII]}$ relation as well as close to the lower SFR and lower $L_{\rm [CII]}$ side of the low-$z$ LIRGs region and the border between data points of the LIRGs and the local star-forming galaxies, as seen in Figure \ref{SFR_vs_LCII}. We plot the SFR range estimated from the SFR$_{\rm UV}$ and the SFR$_{\rm dust}$ upper limit derived by \citet{Ohyama04} and \citet{Carilli13}, respectively. Ly$\alpha$-1 would be a system different from $z\sim6.6$--7 LAEs and possibly similar to low-$z$ LIRGs. Actually, it is in the vicinity of a system BR 1202-0725 comprising a quasar, an SMG and another LAE, Ly$\alpha$-2, all at $z\sim4.7$ \citep[e.g.,][]{Hu96,Ohta00,Iono06,Wagg12b,Carniani13,Carilli13,Williams14}. Note that we do not plot Ly$\alpha$-2 in Figure \ref{SFR_vs_LCII} because, though detected by ALMA, its $L_{\rm [CII]}$ cannot be measured accurately due to the partial truncation at the edge of the ALMA band \citep{Carilli13,Decarli14}. Though upper limit, Ly$\alpha$-1 has an FIR luminosity $L_{\rm FIR} < 3.6 \times 10^{11} L_{\odot}$ \citep{Carilli13}, comparable to those of LIRGs ($L_{\rm FIR} > 10^{11} L_{\odot}$). In addition, Ly$\alpha$-2 has $L_{\rm FIR} \sim 1.7 \times 10^{12} L_{\odot}$ \citep{Carilli13}, even as luminous as ULIRGs ($L_{\rm FIR} > 10^{12} L_{\odot}$). Moreover, both LAEs have Ly$\alpha$ emission broader ($\sim 1300$ km s$^{-1}$) than most LAEs at similar redshifts \citep{Williams14}, and Ly$\alpha$-1 is also spatially extended in the Ly$\alpha$ narrowband image \citep{Hu96}. Several studies have reported that LAEs at $z\sim2$--4.5 detected in FIR continua and having $L_{\rm FIR}$ as luminous as ULIRGs also exhibit extended Ly$\alpha$ emission, and some of them are even classified as LABs showing evidence of interaction within the Ly$\alpha$ cloud. They also tend to be found in dense environments such as protoclusters \citep{Steidel00,Chapman04,Colbert06,Matsuda11,Bridge12,Yang12,Yang14}. Detecting new faint sources, possibly additional LAEs also associated with the quasar and the SMG, \citet{Carniani13} argue that the BR 1202-0725 system might be a massive protocluster in which the high star formation activity is probably triggered by minor mergers or interactions. Hence, Ly$\alpha$-1 is probably a system not like usual LAEs such as IOK-1 but rather possibly a fainter and less extended analog of LIRGs/LABs seen in dense environments. Also, at $z\sim4.7$, the Universe is a factor $\sim 1.5$--1.7 older than the one at $z\sim 6.6$--7. Hence, the difference between Ly$\alpha$-1 and the $z\sim 6.6$--7 LAEs may partly originate from galaxy evolution over cosmic time as well.    

In Figure \ref{SFR_vs_LCII}, we also plot a $z=5.295$ triple LBG system, LBG-1, detected in [CII] but undetected in dust continuum with ALMA and associated with a $z\sim 5.3$ protocluster including a $z=5.2988$ SMG, AzTEC-3 \citep{Capak11,Riechers14}. LBG-1 has an SFR comparable to those of $z\sim 4.7$--7 LAEs in Figure \ref{SFR_vs_LCII}, but its $L_{\rm [CII]}$ is about an order of magnitude higher than the LAEs (except for SDF J132408.3+271543 and SDF J132415.7+273058 due to their shallow $L_{\rm [CII]}$ limits). Moreover, LBG-1 is on the two local galaxy SFR--$L_{\rm [CII]}$ correlations and at the high SFR and high $L_{\rm [CII]}$ edge of the local star-forming galaxy data distribution. According to the analysis of \citet{Riechers14}, though LBG-1 comprises three LBG components and is associated with a protocluster, it does not seem to be a galaxy of extreme type but rather ``typical'', $\sim L_{\rm UV}^*$ star-forming system with relatively low dust content. Furthermore, \citet{Riechers14} compared the SED of LBG-1 including their ALMA dust continuum upper limit with those of local galaxies (mostly the same local galaxy templates we and \citet{Ouchi13} used for the same purpose for IOK-1 and Himiko) and found that LBG-1 is similar to local dwarf/irregular galaxies. Hence, all the information taken together implies that both typical LAEs and LBGs would share SEDs similar to those of local dwarf/irregular galaxies, but LBGs would be more enriched in gas (or higher metallicity) or have different PDR structure/ISM physical state than LAEs. However, to securely confirm this, more observations and sample statistics of typical high redshift LBGs and LAEs are required. Also, note that LBG-1 is not detected in dust continuum, but the ALMA observaton of this object by \citet{Riechers14} is a factor 3--7 shallower than those of IOK-1, Himiko and HCM 6A. Thus, observations of LBGs and LAEs to comparable depth is necessary to fairly compare these two galaxy populations in the context of dust content.    
\begin{figure*}
%\epsscale{1.22}
%\plotone{Figure_LCII_to_LFIR_vs_LFIR_50kmps_40kmps.eps}
%\plotone{Figure_LCII_to_LFIR_vs_LFIR_50kmps_40kmps_LMC.eps}
%\plotone{Figure_LCII_to_LFIR_vs_LFIR_50kmps_40kmps_LMC-2beamsIOK1-11beamsHimiko.eps}
%\plotone{Figure_LCII_to_LFIR_vs_LFIR_50kmps_40kmps_LMC-taperedIOK1-taperedHimiko.eps}
%\plotone{Figure_LCII_to_LFIR_vs_LFIR_50kmps_40kmps_LMC-taperedIOK1-taperedHimiko_Reversed2.eps}
\plotone{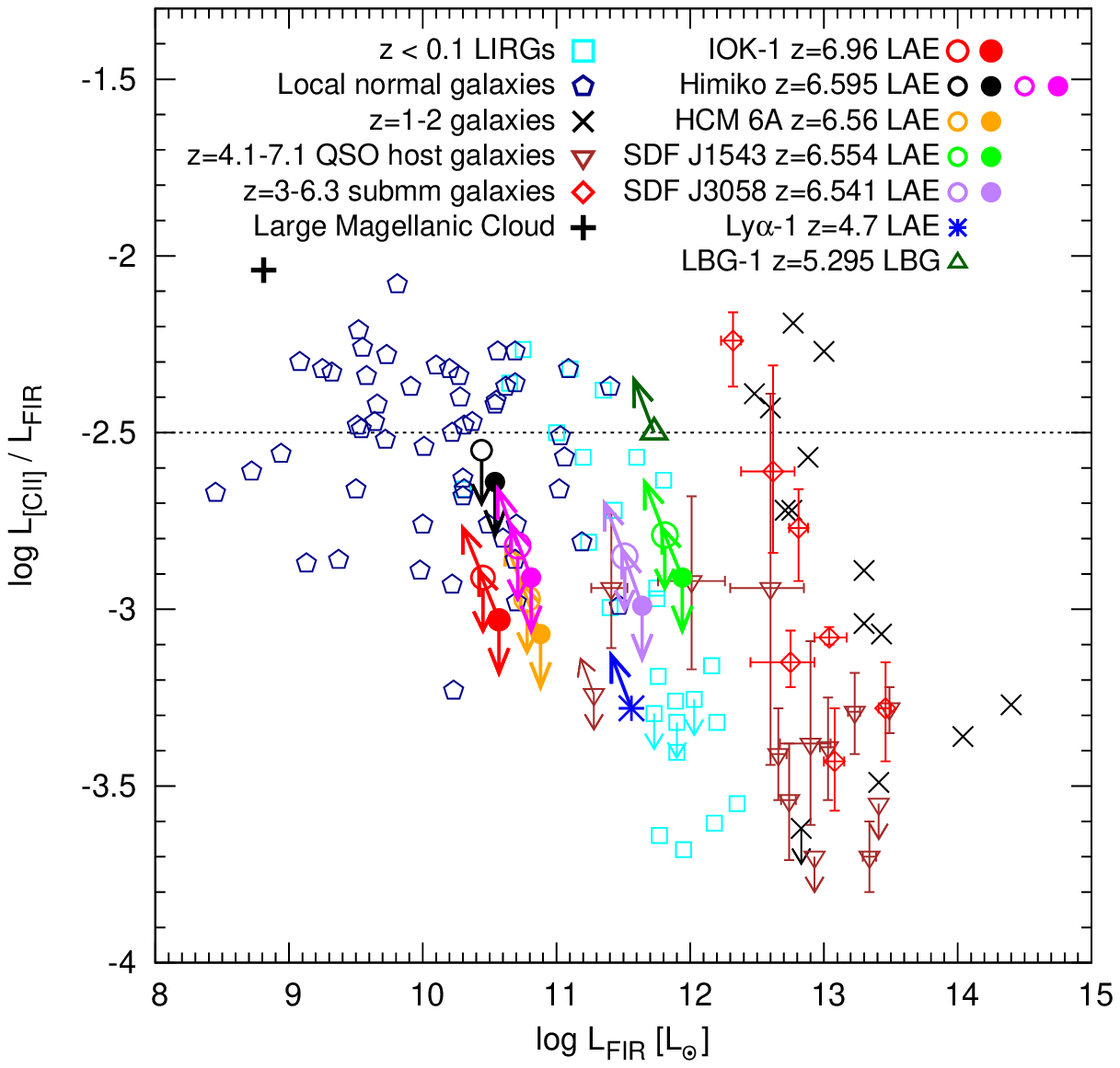}
\caption{Ratio $L_{\rm [CII]}/L_{\rm FIR}$ as a function of FIR luminosity $L_{\rm FIR}$ plotted for different types of objects (See text in \textsection \ref{LCII2FIR_vs_LFIR} for the references from which each data was taken). The horizontal dotted line indicates the average $L_{\rm [CII]}/L_{\rm FIR}$ ratio of the local normal galaxies we calculated from the data. The color coded filled and open circles with arrows indicate $z\sim 6.6$--7 LAE data with $L_{\rm FIR}$ corrected and uncorrected for the CMB effects, respectively (See text in \textsection \ref{LCII2FIR_vs_LFIR} for details). As for Himiko, the magenta filled and open circles are the data based on the CMB-corrected and CMB-uncorrected upper limits on $L_{\rm FIR}$, respectively, at $2.''01 \times 1.''85$ resolution comparable to the size of Himiko in Ly$\alpha$, where we assume that Himiko is not detected in dust continuum at any resolution. The black filled and open circles are the Himiko data based on the CMB-corrected and CMB-uncorrected $L_{\rm FIR}$'s, respectively, at $0.''8 \times 0.''5$ resolution, where we assume that Himiko is detected in dust continuum at 3.4 $\sigma$ significance at this resolution (see the text in \textsection \ref{FIR_Properties} for details). We also plot the data of a $z\simeq 4.7$ LAE, Ly$\alpha$-1, in the quasar $+$ SMG system BRI 1202-0725 \citep{Carilli13}. The arrows of each LAE data (except for the Ly$\alpha$-1 data and the Himiko data shown by the black filled and open circles) indicate the upper limits on the luminosity ratio $L_{\rm [CII]}/L_{\rm FIR}$ at a given $L_{\rm FIR}$ lower than the upper limit on the $L_{\rm FIR}$ of that LAE. \label{LCIIoverLFIR_vs_LFIR}} 
\end{figure*}
%\footnote{http://ned.ipac.caltech.edu/}
%The shaded area shows the possible parameter space for IOK-1 based on our upper limits. Assuming that the FIR-based (extincted) SFR equals the UV-based (unextincted) SFR (i.e., an FIR luminosity that is three times lower than our upper limit) results in the upper limit datapoint shown for the IOK-1. The horizontal dashed line indicates the average value of $L_{\rm [CII]}/L_{\rm FIR}=3 \times 10^{-3}$ for local galaxies. 

\subsubsection{[CII] to FIR Luminosity Ratio \label{LCII2FIR_vs_LFIR}}
Figure \ref{LCIIoverLFIR_vs_LFIR} shows the luminosity ratio $L_{\rm [CII]}/L_{\rm FIR}$ as a function of FIR luminosity $L_{\rm FIR}$ plotted for different types of objects: $z<0.1$ LIRGs \citep{Maiolino09}, local normal galaxies \citep{Malhotra01}\footnote[1]{We converted their FIR fluxes given in a unit of W m$^{-2}$ in their paper to those in $L_{\odot}$ by using the luminosity distances of these galaxies in the cosmology we adopted taken from the NASA/IPAC Extragalactic Database: http://ned.ipac.caltech.edu/}, $z=1$--2 galaxies including starburst-dominated, AGN-dominated and mixed systems \citep{Stacey10}, $z=4.1$--7.1 quasar host galaxies \citep{Pety04,Maiolino05,Iono06,Maiolino09,Wagg10,Venemans12,Wang13,Willott13}, $z=3$--6.34 submm galaxies \citep{Iono06,Cox11,DeBreuck11,Valtchanov11,Walter12a,Riechers13,Rawle14} and a well known dwarf irregular galaxy, Large Magellanic Cloud (LMC) \citep{Rubin09} as well as $z\sim4.7$--7 LAEs, IOK-1 (this study), Himiko \citep{Ouchi13}, HCM 6A \citep{Kanekar13}, SDF J132408.3+271543, SDF J132415.7+273058 \citep{Gonzalez-Lopez14}, Ly$\alpha$-1 in the quasar $+$ SMG system BRI 1202-0725 \citep{Carilli13} and a $z=5.295$ triple LBG system LBG-1 associated with a $z\sim 5.3$ protocluster \citep{Capak11, Riechers14}. The [CII] line is the $^2$P$_{3/2}$ $\rightarrow$ $^2$P$_{1/2}$ fine-structure line that could be emitted by the gas exposed to far-UV radiation in the PDRs, X-ray dominated regions (XDRs) related to an AGN activity and/or HII regions \citep[e.g.,][]{Kaufman99,Meijerink05,Stacey10}. Hence, the fraction of [CII] line luminosity compared to the total FIR luminosity varies with type of object, reflecting its structure of PDR, XDR and/or HII region and physical state of ISM. For example, normal star-forming galaxies in the local Universe tend to show very strong [CII] emission, typically nearly up to an order of magnitude stronger luminosity ratio ($\log L_{\rm [CII]}/L_{\rm FIR} \sim -2.5$ on average) than AGN powered systems such as quasar host galaxies, (U)LIRGs and SMGs. Hence, $L_{\rm [CII]}/L_{\rm FIR}$ versus $L_{\rm FIR}$ diagram can distinguish between different types of objects. 
%The [CII] is the $^2$P$_{3/2}$ $\rightarrow$ $^2$P$_{1/2}$ fine-structure line emitted mainly by the gas exposed to far-UV radiation in the photodissociation regions (PDRs) associated with star-forming activity. Hence, fraction of [CII] line luminosity compared to the total FIR luminosity varies with type of object, reflecting its structure of star-forming region and PDR and physical state of ISM.

In Figure \ref{LCIIoverLFIR_vs_LFIR}, we compare the $z\sim4.7$--7 LAEs with other objects. The color coded filled and open circles with arrows indicate their data with $L_{\rm FIR}$ corrected and uncorrected for the CMB effects, respectively. For IOK-1, Himiko and HCM 6A, we calculated their $L_{\rm FIR}$ limits with and without the CMB effects corrected by integrating from 42.5 $\mu$m to 122.5 $\mu$m the MBBs with $T_{\rm dust}^{z=0}=27.6$ K and $\beta=1.5$ \citep[average of dwarf/irregular galaxies;][]{Skibba11} scaled to the upper limits on the FIR continuum flux density, which we corrected and did not correct for the CMB effects. For Himiko, we plot the data based on $L_{\rm FIR}$ upper limits at $2.''01 \times 1.''85$ resolution (comparable to the size of Himiko in Ly$\alpha$) where we assume that Himiko is not detected in dust continuum at any resolution as well as the data based on the $L_{\rm FIR}$'s at $0.''8 \times 0.''5$ resolution where we assume that Himiko is detected in dust continuum at 3.4 $\sigma$ significance at this resolution (see the text in \textsection \ref{FIR_Properties} for details). For HCM 6A, we used its FIR continuum flux density limit taken from \citet{Kanekar13}. For SDF J132408.3+271543 and SDF J132415.7+273058, we calculated their $L_{\rm FIR}$ limits with and without the CMB effects corrected by integrating from 42.5 $\mu$m to 122.5 $\mu$m the template SED of NGC 6946 \citep{Silva98} scaled to their upper limits on the FIR continuum flux density provided in \citet{Gonzalez-Lopez14}, as they have done so. For each of $z\simeq 6.6$--7 LAEs, we used its $L_{\rm [CII]}$ limit over a channel width we and each author adopted: 40, 36, 100, 50 and 50 km s$^{-1}$ for IOK-1, Himiko, HCM 6A, SDF J132408.3+271543 and SDF J132415.7+273058, respectively \citep{Kanekar13,Gonzalez-Lopez14}. We also plot the data of Ly$\alpha$-1 derived by \citet{Carilli13} with the asterisks. The arrows of each LAE data indicate the upper limits on the luminosity ratio $L_{\rm [CII]}/L_{\rm FIR}$ at a given $L_{\rm FIR}$ lower than the upper limit on the $L_{\rm FIR}$ of that LAE. Comparing the filled and open circles in Figure \ref{LCIIoverLFIR_vs_LFIR}, we notice that omitting to correct for the CMB effects results in underestimation of $L_{\rm FIR}$ and thus overestimation of the ratio $L_{\rm [CII]}/L_{\rm FIR}$ as discussed in \textsection \ref{FIR_Properties}.

As seen in Figure \ref{LCIIoverLFIR_vs_LFIR}, IOK-1, Himiko and HCM 6A are in the region occupied by local normal star-forming galaxies. However, these LAEs are apart from the average point of local normal star-forming galaxies, where $\log L_{\rm FIR} \sim 10.6$ and $\log L_{\rm [CII]}/L_{\rm FIR} \sim -2.5$, and IOK-1 and HCM 6A are even at the lower $L_{\rm [CII]}/L_{\rm FIR}$ edge of the local normal star-forming galaxy region. Meanwhile, SDF J132408.3+271543 and SDF J132415.7+273058 are in the low redshift LIRG region, probably because their upper limits on $L_{\rm [CII]}$ and $L_{\rm FIR}$ are an order of magnitude shallower than those of other LAEs. Unless we observe them to much deeper limits, we cannot tell if they are also located in the similar positions as the other $z\sim6.6$--7 LAEs in Figure \ref{LCIIoverLFIR_vs_LFIR}. 

We also notice that the possible location of IOK-1 and HCM 6A in Figure \ref{LCIIoverLFIR_vs_LFIR} includes the data point of LMC, the most nearby dwarf irregular galaxy. This might be consistent with a picture that high redshift LAEs could be similar to local dwarf/irregular galaxies, as inferred from the comparison of the SED of IOK-1 with those of local dwarfs/irregulars in Figure \ref{SED_IOK-1}. However, this dose not necessary mean that the similarity in dust continuum SEDs guarantees the similarity in $L_{\rm [CII]}/L_{\rm FIR}$ ratios, as we have not detected [CII] lines and dust continua of IOK-1 and HCM 6A and uncertainties in their $L_{\rm [CII]}/L_{\rm FIR}$ are large. In fact, while the possible location of Himiko in Figure \ref{LCIIoverLFIR_vs_LFIR} in the case that we assume a non-detection in dust continuum as an extended source (the magenta filled circle) includes the data point of LMC, the $L_{\rm [CII]}/L_{\rm FIR}$ upper limit of Himiko in the case that we assume a detection in dust continuum (the black filled circle) is not consistent with LMC, although both cases (i.e., the $L_{\rm FIR}$ upper limit and ``detected'' $L_{\rm FIR}$ value) are consistent with the dust continuum SEDs of the dwarf/irregular galaxies in Figure 5 of \citet{Ouchi13}. Thus, it seems that the similarity in dust continuum SED does not necessarily mean similar $L_{\rm [CII]}/L_{\rm FIR}$ ratios. Also, we neither have the dust continuum SED data of LMC that can be compared with those of IOK-1 and Himiko nor $L_{\rm [CII]}/L_{\rm FIR}$ ratio data of dwarf/irregular galaxies whose dust continuum SEDs were compared with those of IOK-1 and Himiko in Figure \ref{SED_IOK-1} in this paper and Figure 5 in \citet{Ouchi13}. Thus, at this moment we cannot tell if high redshift LAEs are similar to local dwarf/galaxies in dust continuum SED and $L_{\rm [CII]}/L_{\rm FIR}$ ratio simultaneously. To reveal this, we need statistically large number of dwarf/irregular galaxy data whose dust continuum SEDs and $L_{\rm [CII]}/L_{\rm FIR}$ ratios are available at the same time.

On the other hand, Ly$\alpha$-1 is in the low redshift LIRG region or at the border between the regions of local normal star-forming galaxies and low redshift LIRGs in Figure \ref{LCIIoverLFIR_vs_LFIR}. This could be because it is associated with the special dense environment consisting of a quasar, an SMG and other LAE and likely a system different from common LAEs but more similar to LIRGs as discussed in \textsection \ref{CII_SFR}. 

As for LBG-1 in Figure \ref{LCIIoverLFIR_vs_LFIR}, it has a lower $L_{\rm FIR}$ and/or a higher $L_{\rm [CII]}/L_{\rm FIR}$ ratio than quasar host galaxies, submm galaxies and $z=1$--2 galaxies and is clearly separate from them. LBG-1 is rather close to the locations of low redshift LIRGs and the high $L_{\rm FIR}$ edge of local normal galaxies but far apart from the $z\sim 4.7$--7 LAEs. It should be noted that as we mentioned earlier, the ALMA observaton of LBG-1 by \citet{Riechers14} is a factor 3--7 shallower than those of IOK-1, Himiko and HCM 6A. Thus, we cannot tell how different the $z\sim 4.7$--7 LAEs and LBG-1 are in $L_{\rm FIR}$ and $L_{\rm [CII]}/L_{\rm FIR}$ ratio from the current data alone. However, Figure \ref{LCIIoverLFIR_vs_LFIR} implies that $z\sim 4.7$--7 LAEs and LBG-1 cannot have similar $L_{\rm FIR}$'s and similar $L_{\rm [CII]}/L_{\rm FIR}$ ratios at the same time, because the [CII] line is detected from LBG-1 at the $L_{\rm [CII]}$ about an order of magnitude higher than the $L_{\rm [CII]}$ or $L_{\rm [CII]}$ limits of $z\sim 4.7$--7 LAEs (see Figure \ref{SFR_vs_LCII}). In any case, observations of more LBGs and LAEs to comparable depth is necessary to fairly compare LAEs and LBGs in the context of $L_{\rm FIR}$ and $L_{\rm [CII]}/L_{\rm FIR}$ ratio. 

After all, the locations of the $z\sim4.7$--7 LAEs in the $L_{\rm [CII]}/L_{\rm FIR}$--$L_{\rm FIR}$ diagram in Figure \ref{LCIIoverLFIR_vs_LFIR} are consistent with what is expected from the locations of these LAEs in the SFR--$L_{\rm [CII]}$ diagram in Figure \ref{SFR_vs_LCII}. These two diagrams together suggest that the $z\sim 6.6$--7 LAEs, IOK-1, Himiko and HCM 6A are similar to but more likely lower than local normal star-forming galaxies in $L_{\rm [CII]}$, $L_{\rm FIR}$ and $L_{\rm [CII]}/L_{\rm FIR}$ ratio. This implies that these LAEs have either lower enriched gas (or metallicity) and dust contents or different PDR structures/physical states of ISM than local star-forming galaxies. Also, the two diagrams imply that the $z\sim 4.7$ LAE, Ly$\alpha$-1, could be a system similar to low redshift LIRGs. 

%\subsubsection{Atomic Masses of High-$z$ LAEs}

%%%%%%%%%%%%%%%%%%%%%%%%%%%%%%%%%%%%%%%%% UNCOMMENT %%%%%%%%%%%%%%%%%%%%%%%%%%%%%%%%%%%%%%%%%
%\subsection{Star Formation Rate}
%The Ly$\alpha$ emission is attenuated by neutral hydrogen in intergalactic medium and dust, while the UV continuum by only dust. Therefore, the ratios, SFR$_{{\rm Ly}\alpha}$/SFR$_{\rm total} \sim XX$ and SFR$_{\rm UV}$/SFR$_{\rm total} \sim XX$, correspond to the escape fraction of Ly$\alpha$ photons $f_{\rm esc}^{{\rm Ly}\alpha}$ and dust extinction $A_{\rm UV}$, respectively, and tell us how much fraction of SFR is obscured or unobscured. Now that we know the $A_{\rm UV}$ and redshift ($z = 6.96$) of IOK-1, we can constrain the age and stellar mass $M_*$ of IOK-1 more accurately than our previous work (Ota et al.~2010b). Fixing the redshift and dust attenuation and fitting the optical to mid-infrared photometry of IOK-1 (Ota et al.~2010b; Cai et al.~2011) to stellar population synthesis models (Bruzual \& Charlot~2003), we estimate that this galaxy is XX million years old and has a stellar mass of $M_* \sim XX \times 10^{X} M_{\odot}$.
%%%%%%%%%%%%%%%%%%%%%%%%%%%%%%%%%%%%%%%%% UNCOMMENT %%%%%%%%%%%%%%%%%%%%%%%%%%%%%%%%%%%%%%%%%

\section{Summary and Conclusion}
We have conducted ALMA 158 $\mu$m [CII] line and FIR continuum observation of a $z=6.96$ LAE, IOK-1, a normally star-forming galaxy in the epoch of cosmic reionization, reaching line and continuum sensitivities of $\sigma_{\rm line} =$ 240 $\mu$Jy beam$^{-1}$ (over a channel width of 40 km s$^{-1}$) and $\sigma_{\rm cont} =$ 21 $\mu$Jy beam$^{-1}$ at a resolution of $1.''5 \times 1.''2$, comaprable of the size IOK-1 in Ly$\alpha$. 

We did not detect the FIR continuum. We constructed the UV--FIR SED of IOK-1 from its Subaru/Suprime-Cam, {\it HST}/WFC3, {\it Spitzer}/IRAC, {\it Herschel}/SPIRE and our ALMA data ($3\sigma$ upper limit of $S_{\rm cont}^{\rm obs} = 63$ $\mu$Jy) and compared it with template SEDs of several types of local galaxies whose FIR continua were both corrected and uncorrected for the CMB effects. We found that IOK-1 is similar to local dwarf and irregular galaxies rather than highly obscured/dusty systems. Also, modeling the dust continuum SED of IOK-1 as the MBB with the dust temperature and the emissivity index typical of dwarf galaxies and scaling it to the CMB-corrected ALMA FIR continuum flux upper limit, we estimated the upper limits on the dust mass, total FIR luminosity ($\lambda=42.5$ -- 122.5 $\mu$m), total IR luminosity ($\lambda=8$ $\mu$m -- 1 mm) and dust-obscured SFR of IOK-1 to be $M_{\rm dust} < 6.4 \times 10^7 M_{\odot}$, $L_{\rm FIR} < 3.7\times 10^{10} L_{\odot}$, $L_{\rm IR} < 5.7 \times 10^{10} L_{\odot}$ and SFR$_{\rm dust} < 10.0 M_{\odot}$ yr$^{-1}$, respectively. Since total SFR of IOK-1 is SFR$_{\rm total} =$ SFR$_{\rm UV}$ $+$ SFR$_{\rm dust} < 33.9$ $M_{\odot}$ yr$^{-1}$, we estimate that $<29$\% of star formation in IOK-1 is obscured by dust. Moreover, we found that if we do not correct for the effects of CMB on FIR continuum flux measurement, we underestimate the physical quantities related to dust, $M_{\rm dust}$, $L_{\rm FIR}$, $L_{\rm IR}$ and SFR$_{\rm dust}$, by 24\%. 

On the other hand, we found a $3\sigma$ weak signal in the [CII] image and spectrum at the position of IOK-1 and at 238.805 GHz, corresponding to the redshift slightly lower than the one previously estimated from Ly$\alpha$ emission. This is consistent with the idea that redshift estimated from Ly$\alpha$ was overestimated due to the absorption by IGM. Since the signal is marginal for securely concluding that it is indeed the [CII] line, we have treated it as non-detection and investigated the properties of IOK-1 related to [CII] line using the $3\sigma$ upper limit of $S_{\rm line}^{\rm obs} = 720$ $\mu$Jy (over a channel width of 40 km s$^{-1}$). This corresponds to the [CII] luminosity of $L_{\rm [CII]} < 3.4 \times 10^7 L_{\odot}$. In the SFR--$L_{\rm [CII]}$ and the $L_{\rm [CII]}/L_{\rm FIR}$--$L_{\rm FIR}$ planes, we compared the locations of IOK-1 as well as other $z\sim4.7$--6.6 LAEs previously observed in the [CII] line and FIR continuum with those of other objects including local to low redshift normal star-forming galaxies, low redshift LIRGs, low to high redshift quasar host galaxies and SMGs and a high redshift LBG. We found that $z\sim 6.6$--7 LAEs at reionization epoch are systems different from LIRGs, quasar hosts, SMGs and the LBG but similar to or even lower than local star-forming galaxies in $L_{\rm [CII]}$, $L_{\rm FIR}$ and $L_{\rm [CII]}/L_{\rm FIR}$ ratio. This implies that these LAEs have either lower enriched gas (or metallicity) and dust or different PDR structures/physical states of ISM than those of local normal star-forming galaxies.

The present study is based on only one $z\simeq7$ LAE (as well as a few other $z\sim4.7$--6.6 LAEs previously observed in [CII] and FIR continua) and results obtained from the analysis of them might not be representative of the [CII] line and FIR continuum properties of the general population of LAEs at the epoch of reionization. Fortunately, the number of antennas has been increasing for ALMA as its construction proceeds towards full capability, substantially improving the sensitivity and resolution. This will soon allow us to investigate [CII] lines and FIR continua of the statistically larger number of high redshift star-forming galaxies at a broader redshift range lower and higher than $z \sim 6$ (epoch of the end of reionization) in reasonable integration times. Such observations will give us more general insight into the obscured star formation, physical properties, ISM state and even dynamics of high redshift star-forming galaxies and will shed light on their relationship with reionization and other galaxy populations.


%% Included in this acknowledgments section are examples of the
%% AASTeX hypertext markup commands. Use \url without the optional [HREF]
%% argument when you want to print the url directly in the text. Otherwise,
%% use either \url or \anchor, with the HREF as the first argument and the
%% text to be printed in the second.

\acknowledgments
This paper makes use of the following ALMA data: ADS/JAO.ALMA\# 2011.0.00767.S and 2011.0.00115.S. ALMA is a partnership of ESO (representing its member states), NSF (USA) and NINS (Japan), together with NRC (Canada) and NSC and ASIAA (Taiwan), in cooperation with the Republic of Chile. The Joint ALMA Observatory is operated by ESO, AUI/NRAO and NAOJ. We thank Rob J. Ivison, Steve A. Eales and Matthew W.L. Smith for kindly providing us with the {\it Herschel}-ATLAS images around IOK-1. The {\it Herschel}-ATLAS is a project with {\it Herschel}, which is an ESA space observatory with science instruments provided by European-led Principal Investigator consortia and with important participation from NASA. The H-ATLAS website is http://www.h-atlas.org/. This work was partly supported by the Grant-in-Aid for the Global COE Program "The Next Generation of Physics, Spun from Universality and Emergence" from the Ministry of Education, Culture, Sports, Science and Technology (MEXT) of Japan. We thank Takatoshi Shibuya, Steven Furlanetto and Daniel Schaerer for their comments on the paper. K.O. acknowledges the Kavli Institute Fellowship at the Kavli Institute for Cosmology in the University of Cambridge supported by the Kavli Foundation. K.O. was supported by the ALMA Japan Research Grant of NAOJ Chile Observatory, NAOJ-ALMA-0017. We thank our referee for useful comments that helped us to improve this paper.
%This work is based in part on data collected at the Subaru Telescope, which is operated by the National Astronomical Observatory of Japan (NAOJ).
%B.H. was supported with a fellowship from the Japan Society for the Promotion of Science. 
%KO acknowledges the Kavli Institute Fellowship support from the Kavli Foundation.


%% in parentheses, after the keyword, but they will not be verified.

{\it Facilities:} \facility{ALMA, Subaru (Suprime-Cam), {\it HST} (WFC3), {\it Spitzer} (IRAC), {\it Herschel} (SPIRE)}


\begin{thebibliography}{ }
\bibitem[Beelen et al.(2006)]{Beelen06} Beelen, A., Cox, P., Benford, D. J., et al. 2006, \apj, 642, 694
\bibitem[Boone et al.(2007)]{Boone07} Boone, F., Schaerer, D., Pell\'o, R., Combes, F., \& Egami, E. 2007, \aap, 475, 513
\bibitem[Boselli et al.(2002)]{Boselli02} Boselli, A., Gavazzi, G., Lequeux, J., \& Pierini, D. 2002, \aap, 385, 454
\bibitem[Bouwens et al.(2012)]{Bouwens12} Bouwens, R. J., Illingworth, G. D., Oesch, P. A., et al. 2012, \apjl, 752, L5
\bibitem[Bridge et al.(2012)]{Bridge12} Bridge, C. R., Blain, A., Borys, C. J. K., et al. 2013, \apj, 769, 91
\bibitem[Cai et al.(2011)]{Cai11} Cai, Z., Fan, X., Jiang, L., et al. 2011, \apjl, 736, L28
\bibitem[Capak et al.(2011)]{Capak11} Capak, P. L., Riechers, D., Scoville, N. Z., et al. 2011, \nat, 470, 233
\bibitem[Carilli et al.(2013)]{Carilli13} Carilli, C. L., Riechers, D., Walter, F., et al. 2013, \apj, 763, 120
\bibitem[Carniani et al.(2013)]{Carniani13} Carniani, S., Marconi, A., Biggs, A., et al. 2013, \aap 559, A29
\bibitem[Chabrier (2003)]{Chabrier03} Chabrier, G. 2003, PASP, 115, 763
\bibitem[Chapman et al.(2004)]{Chapman04} Chapman, S. C., Scott, D., Windhorst, R. A., et al. 2004, \apj, 606, 85
\bibitem[Colbert et al.(1999)]{Colbert99} Colbert, J. W., Malkan, M. A., Clegg, P. E., et al. 1999, \apj, 511, 721 
\bibitem[Colbert et al.(2006)]{Colbert06} Colbert, J. W., Teplitz, H., Francis, P., et al. 2006, \apjl, 637, L89 
\bibitem[Cox et al.(2011)]{Cox11} Cox, P., Krips, M., Neri, R., et al. 2011, \apj, 740, 63
\bibitem[da Cunha et al.(2013)]{daCunha13} da Cunha, E., Groves, B., Walter, F., et al. 2013, \apj, 766, 13 
\bibitem[Dale et al.(2007)]{Dale07} Dale, D. A., Gil de Paz, A., Gordon, K. D., et al. 2007, \apj, 655, 863 
\bibitem[De Breuck et al.(2011)]{DeBreuck11} De Breuck, C., Maiolino, R., Caselli, P., et al. 2011, \aap, 530, L8
\bibitem[Decarli et al.(2014)]{Decarli14} Decarli, R., Walter, F., Carilli, C., et al. 2014, \apjl, 782, L17
\bibitem[De Looze et al.(2011)]{DeLooze11} De Looze, I., Baes, M., Bendo, G. J., Cortese, L., \& Fritz, J. 2011, \mnras, 416, 2712 
\bibitem[Dunlop (2013)]{Dunlop13} Dunlop, J. S. 2013, in The First Galaxies (Astrophysics and Space Science Library, Vol. 396; Berlin: Springer-Verlag), 223
\bibitem[Dunne et al.(2000)]{Dunne00} Dunne, L., Eales, S., Edmunds, M., et al. 2000, \mnras, 315, 115 
\bibitem[Dwek et al.(2011)]{Dwek11} Dwek, E., Staguhn, J. G., Arendt, R. G., et al. 2011, \apj, 738, 36 
\bibitem[Eales et al.(2010)]{Eales10} Eales, S., Dunne, L., Clements, D., et al. 2010, PASP, 122, 499
\bibitem[Elbaz et al.(2007)]{Elbaz07} Elbaz, D., Daddi, E., Le Borgne, D. et al. 2007, \aap, 468, 33
\bibitem[Finkelstein et al.(2012)]{Finkelstein12} Finkelstein, S. L., Papovich, C., Ryan, R. E., Jr., et al. 2012, \apj, 758, 93 
\bibitem[Finkelstein et al.(2013)]{Finkelstein13} Finkelstein, S. L., Papovich, C., Dickinson, M., et al. 2013, \nat, 502, 524 
\bibitem[Gonz\'alez-L\'opez et al.(2014)]{Gonzalez-Lopez14} Gonz\'alez-L\'opez, J., Riechers, D. A., Decarli, R., et al. 2014, \apj, 784, 99 
%\bibitem[Hinshaw et al.(2013)]{Hinshaw13} Hinshaw, G., Larson, D., Komatsu, E., et al. 2013, \apjs, 208, 19  
\bibitem[Hu et al.(1996)]{Hu96} Hu, E. M., McMahon, R. G., \& Egami, E. 1996, \apjl, 459, L53 
\bibitem[Hu et al.(2002a)]{Hu02a} Hu, E. M., Cowie, L. L., McMahon, R. G., et al. 2002a, \apjl, 568, L75 
\bibitem[Hu et al.(2002b)]{Hu02b} Hu, E. M., Cowie, L. L., McMahon, R. G., et al. 2002b, \apjl, 576, L99 
\bibitem[Hu et al.(2010)]{Hu10} Hu, E. M., Cowie, L. L., Barger, A. J., et al. 2010, \apj, 725, 394 
\bibitem[Iono et al.(2006)]{Iono06} Iono, D., Yun, M. S., Elvis, M., et al. 2006, \apjl, 645, L97 
\bibitem[Ivison et al.(2010)]{Ivison10} Ivison, R. J., Swinbank, A. M., Swinyard, B., et al. 2010, \aap, 518, L35
\bibitem[Iye et al.(2006)]{Iye06} Iye, M., Ota, K., Kashikawa, N., et al. 2006, \nat, 443, 186
\bibitem[Jiang et al.(2013)]{Jiang13} Jiang, L., Egami, E., Mechtley, M., et al. 2013, \apj, 772, 99
\bibitem[Kanekar et al.(2013)]{Kanekar13} Kanekar, N., Wagg, J., Ram Chary, R., \& Carilli, C. 2013, \apj, 771, 20
\bibitem[Kashikawa et al.(2011)]{Kashikawa11} Kashikawa, N., Shimasaku, K., Matsuda, Y., et al. 2011, \apj, 734, 119 
\bibitem[Kaufman et al.(1999)]{Kaufman99} Kaufman, M. J., Wolfire, M. G., Hollenbach, D. J., \& Luhman, M. L. 1999, \apj, 527, 795
\bibitem[Kennicutt(1998)]{Kennicutt98} Kennicutt, R. C. 1998, \araa, 36, 189 
\bibitem[Konno et al.(2014)]{Konno14} Konno, A., Ouchi, M., Ono, Y., et al. 2014, \apj, submitted (arXiv1404.6066)
\bibitem[Kroupa (2001)]{Kroupa01} Kroupa, P. 2001, \mnras, 322, 231
\bibitem[Maiolino et al.(2005)]{Maiolino05} Maiolino, R., Cox, P., Caselli, P., et al. 2005, \aap, 440, L51 
\bibitem[Maiolino et al.(2009)]{Maiolino09} Maiolino, R., Caselli, P., Nagao, T., et al. 2009, \aap, 500, L1 
\bibitem[Malhotra et al.(2001)]{Malhotra01} Malhotra, S., Kaufman, M. J., Hollenbach, D., et al. 2001, \apj, 561, 766 
\bibitem[Matsuda et al.(2011)]{Matsuda11} Matsuda, Y., Yamada, T., Hayashino, T., et al. 2011, \mnras, 410, L13 
\bibitem[McLinden et al.(2011)]{McLinden11} McLinden, E. M., Finkelstein, S. L., Rhoads, J. E. et al. 2010, \apj, 730, 136
\bibitem[Meijerink \& Spaans(2005)]{Meijerink05} Meijerink, R., \& Spaans, M. 2005, \aap, 436, 397
\bibitem[Mentuch Cooper et al.(2012)]{MentuchCooper12} Mentuch Cooper, E., Wilson, C. D., Foyle, K., et al. 2012, \apj, 755, 165 
\bibitem[Momose et al.(2014)]{Momose14} Momose, R., Ouchi, M., Nakajima, K., et al. 2014, \mnras, 442, 110  
\bibitem[Ohta et al.(2000)]{Ohta00} Ohta, K., Matsumoto, T., Maihara, T., et al. 2000, \pasj, 52, 557 
\bibitem[Ohyama et al.(2004)]{Ohyama04} Ohyama, Y., Taniguchi, Y., \& Shioya, Y. 2004, \aj, 128, 2704 
\bibitem[Ono et al.(2010)]{Ono10} Ono, Y., Ouchi, M., Shimasaku, K., et al. 2010, \apj, 724, 1524
\bibitem[Ono et al.(2012)]{Ono12} Ono, Y., Ouchi, M., Mobasher, B., et al. 2012, \apj, 744, 83 
\bibitem[Ota et al.(2008)]{Ota08} Ota, K., Iye, M., Kashikawa, N., et al. 2008, \apj ,677, 12 
\bibitem[Ota et al.(2010a)]{Ota10a} Ota, K., Iye, M., Kashikawa, N., et al. 2010a, \apj, 722, 803
\bibitem[Ota et al.(2010b)]{Ota10b} Ota, K., Ly, C., Malkan, M. A., et al. 2010b, \pasj, 62, 1167
\bibitem[Ouchi et al.(2009a)]{Ouchi09a} Ouchi, M., Ono, Y., Egami, E., et al. 2009a, \apj, 696, 1164
\bibitem[Ouchi et al.(2009b)]{Ouchi09b} Ouchi, M., Mobasher, B., Shimasaku, K., et al. 2009b, \apj, 706, 1136
\bibitem[Ouchi et al.(2010)]{Ouchi10} Ouchi, M., Shimasaku, K., Furusawa, H., et al. 2010, \apj, 723, 869
\bibitem[Ouchi et al.(2013)]{Ouchi13} Ouchi, M., Ellis, R., Ono, Y., et al. 2013, \apj, 778, 102 
\bibitem[Pascale et al.(2011)]{Pascale11} Pascale, E., Auld, R., Dariush, A., et al. 2011, \mnras, 415, 911 
\bibitem[Pettini et al.(2001)]{Pettini01} Pettini, M., Shapley, A. E., Steidel, C. C., et al. 2001, \apj, 554, 981
\bibitem[Pety et al.(2004)]{Pety04} Pety, J., Beelen, A., Cox, P., et al. 2004, \aap, 428, L21
%\bibitem[Planck Collaboration (2013)]{Planck13} Planck Collaboration 2013, \aap, submitted (arXiv1303.5076)
\bibitem[Rangwala et al.(2011)]{Rangwala11} Rangwala, N., Maloney, P. R., Glenn, J., et al. 2011, \apj, 743, 94
\bibitem[Rawle et al.(2014)]{Rawle14} Rawle, T. D., Egami, E., Bussmann, R. S., et al. 2014, \apj, 783, 59
\bibitem[Riechers et al.(2010)]{Riechers10} Riechers, D. A., Capak, P. L., Carilli, C. L., et al. 2010, \apjl, 720, L131
\bibitem[Riechers et al.(2013)]{Riechers13} Riechers, D. A., Bradford, C. M., Clements, D. L., et al. 2013, \nat, 496, 329
\bibitem[Riechers et al.(2014)]{Riechers14} Riechers, D. A., Carilli, C. L., Capak, P. L., et al. 2014, \apj, submitted (arXiv1404.7159)
%\bibitem[Robertson et al.(2010)]{Robertson10} Robertson, B. E., Ellis, R. S., Dunlop, J. S., McLure, R. J., \& Stark, D. P. 2010, \nat, 468, 49
\bibitem[Robertson et al.(2013)]{Robertson13} Robertson, B. E., Furlanetto, S. R., Schneider, E., et al. 2013, \apj, 768, 71
\bibitem[Rubin et al.(2009)]{Rubin09} Rubin, D., Hony, S., Madden, S. C., et al. 2009, \aap, 494, 647
\bibitem[Salim et al.(2007)]{Salim07} Salim, S., Rich, R. M.,  Charlot, S. et al. 2007, \apjs, 173, 267
\bibitem[Salpeter(1955)]{Salpeter55} Salpeter, E. E. 1955, \apj, 121, 161 
\bibitem[Shapley et al.(2003)]{Shapley03} Shapley, A. E., Steidel, C. C., Pettini, M., \& Adelberger, K. L. 2003, \apj, 588,
65
\bibitem[Shibuya et al.(2012)]{Shibuya12} Shibuya, T., Kashikawa, N., Ota, K., et al. 2012, \apj, 752, 114
\bibitem[Shibuya et al.(2014)]{Shibuya14} Shibuya, T., Ouchi, M., Nakajima, K., et al. 2014, \apj, 788, 74
\bibitem[Silva et al.(1998)]{Silva98} Silva, L., Granato, G. L., Bressan, A., \& Danese, L. 1998, \apj, 509, 103 
\bibitem[Skibba et al.(2011)]{Skibba11} Skibba, R. A., Engelbracht, C. W., Dale, D., et al. 2011,\apj, 738, 89 
\bibitem[Stacey et al.(2010)]{Stacey10} Stacey, G. J., Hailey-Dunsheath, S., Ferkinhoff, C., et al. 2010, \apj, 724, 957
\bibitem[Steidel et al.(2000)]{Steidel00} Steidel, C. C., Adelberger, K. L., Shapley, A. E., et al. 2000, \apj, 532, 170
\bibitem[Steidel et al.(2010)]{Steidel10} Steidel, C. C., Erb, D. K., Shapley, A. E., et al. 2010, \apj, 717, 289
\bibitem[Taniguchi et al.(2005)]{Taniguchi05} Taniguchi, Y., Ajiki, M., Nagao, T., et al. 2005, \pasj, 57, 165
\bibitem[Vallini et al.(2013)]{Vallini13} Vallini, L., Gallerani, S., Ferrara, A., \& Baek, S. 2013, \mnras, 433, 1567
\bibitem[Valtchanov et al.(2011)]{Valtchanov11} Valtchanov, I., Virdee, J., Ivison, R. J., et al. 2011, \mnras, 415, 3473 
\bibitem[Venemans et al.(2012)]{Venemans12} Venemans, B. P., McMahon, R. G., Walter, F., et al. 2012, \apjl, 751, L25
\bibitem[Wagg et al.(2009)]{Wagg09} Wagg, J., Kanekar, N., \& Carilli, C. L. 2009 \apjl, 697, L33 
\bibitem[Wagg et al.(2010)]{Wagg10} Wagg, J., Carilli, C. L., Wilner, D. J., et al. 2010, \aap, 519, L1 
\bibitem[Wagg et al.(2012a)]{Wagg12a} Wagg, J., \& Kanekar, N. 2012a, \apjl, 751, L24
\bibitem[Wagg et al.(2012b)]{Wagg12b} Wagg, J., Wiklind, T., Carilli, C. L., et al. 2012b, \apjl, 752, L30
\bibitem[Walter et al.(2003)]{Walter03} Walter, F., Bertoldi, F., Carilli, C., et al. 2003, \nat, 424, 406
%\bibitem[Walter et al.(2004)]{Walter04} Walter, F., Carilli, C. L., Bertoldi, F., et al. 2004, \apjl, 615, L17 
\bibitem[Walter et al.(2009)]{Walter09} Walter, F., Riechers, D., Cox, P., et al. 2009, \nat, 457, 699
\bibitem[Walter et al.(2012a)]{Walter12a} Walter, F., Decarli, R., Carilli, C., et al. 2012a, \nat, 486, 233
\bibitem[Walter et al.(2012b)]{Walter12b} Walter, F., Decarli, R., Carilli, C., et al. 2012b, \apj, 752, 93
\bibitem[Wang et al.(2010)]{Wang10} Wang, R., Wagg, J., Carilli, C., et al. 2010, \apjl, 739, L34 
\bibitem[Wang et al.(2013)]{Wang13} Wang, R., Wagg, J., Carilli, C. L., et al. 2013, \apj, 773, 44 
\bibitem[Williams et al.(2014)]{Williams14} Williams, R. J., Wagg, J., Maiolino, R., et al. 2014, \mnras, 439, 2096 
\bibitem[Willott et al.(2013)]{Willott13} Willott, C. J., Omont, A., \& Bergeron, J. 2013, \apj, 770, 13
\bibitem[Yang et al.(2012)]{Yang12} Yang, Y., Decarli, R., Dannerbauer, H., et al. 2012, \apj, 744, 178
\bibitem[Yang et al.(2014)]{Yang14} Yang, Y., Walter, F., Decarli, R., Bertoldi, F., Weiss, A., Dey, A., Prescott, M. K. M., \& Badescu, T. 2014, \apj, 784, 171
\end{thebibliography}
\end{document}